\RequirePackage{fix-cm} 
\PassOptionsToPackage{pdfpagelabels=false}{hyperref} 
\documentclass{svjour3}                     % onecolumn (standard format)
\smartqed  % flush right qed marks, e.g. at end of proof
\usepackage{graphicx}
\usepackage{amssymb}
\usepackage{hyperref}
\usepackage[round,authoryear]{natbib}
\usepackage{latexsym}

\journalname{Stat Papers (2019) 60:1677–1698}

\begin{document}

\title{Detecting a Structural Change in Functional Time Series Using Local Wilcoxon Statistic
}
% Grants or other notes about the article that should go on the front
% page should be placed within the \thanks{} command in the title
% (and the %-sign in front of \thanks{} should be deleted)
%
% General acknowledgments should be placed at the end of the article.
\author{Daniel Kosiorowski \and Jerzy P. Rydlewski \and Ma\l gorzata Snarska
}

\institute{ Daniel Kosiorowski \at
              Department of Statistics, Cracow University of Economics, Krak\'ow, Poland\\                           \email{daniel.kosiorowski@uek.krakow.pl}
           \and
           Jerzy P. Rydlewski (corresponding author) \at
              AGH University of Science and Technology, Faculty of Applied Mathematics, al. A. Mickiewicza 30, 30-059 Krak\'ow, Poland\\
              \email{ry@agh.edu.pl}\\
           tel:(+48) 12 617 31 68, fax: (+48) 12 617 31 65
              \and
              Ma\l gorzata Snarska \at
              Department of Financial Markets, Cracow University of Economics, Krak\'ow, Poland \\            \email{malgorzata.snarska@uek.krakow.pl}
}

\date{Received: 10 May 2016 / Revised: 9 February 2017 / Published online: 28 February 2017/Published: 11 October 2019}

\maketitle

\begin{abstract}
Functional data analysis (FDA) (\citet{Ra,RS}) is a  part of modern multivariate statistics that analyses data providing information about curves, surfaces or anything else varying over a certain continuum.
In economics and empirical finance we often have to deal with time series of functional data, where we cannot easily decide, whether they are to be considered as homogeneous or heterogeneous. 
At present a discussion on adequate tests of homogenity for functional data is carried (see e.g. \citet{Flores}).
We propose a novel statistic for detetecting a structural change in functional time series based on a local Wilcoxon statistic induced by a local depth function proposed in \citet{Pand}.
\keywords{Functional Data Analysis\and Local Depth\and Functional Depth \and Detecting Structural Change \and Heterogenity\and Wilcoxon Test}
\subclass{62G30 \and 62-07 \and 62G35 \and 62P20 }
\end{abstract}

\section{Introduction}
\label{intro}
There are many objects in economics taking form of a function of certain continuum. We mean here utility curves, yield curves, electricity demand trajectories during day and night, time series of concentration of dangerous particulates in atmosphere, the Internet traffic intensity within a day (see fig.1 and fig. 2).
Very often economic phenomena are observed through a certain number of non-homogenous components, i.e. they exhibit multimodality. The phenomena globally seem similar, but locally differ significantly.
Global methods of populations comparison (visual, inferential, descriptive) using popular centrality measures i.e. mean or median may be misleading. Further problems occur when functional outliers (defined e.g. with respect to functional boxplot) are present in  the data set. Due to the lack of reliable economic theory on data generating processes, which are used for describing economic phenomena, functional generalizations of well known statistical procedures (e.g. ANOVA) are inefficient (see \citet{HK}).
Suppose for a while that an economic system in each period of time is described by a certain number of functions (e.g. individual demand and supply curves or investment strategies).
The characteristics of this dynamic system are observed as a multiregime functional time series where heterogenity is related to change in probability distribution over the considered space of functions.
Our aim is to detect a structural change related to local differences between populations or between populations' characteristics in two or more periods of study (e.g. before and after financial crash).\\
In one dimensional case it is known that Wilcoxon ranks sum test correctly detects differences in location for a rich class of populations \citep{RAND}.
Note, that for many economic phenomena described by means of certain curves (e.g. yield curves, utility curves, dangerous particles in atmosphere concentration curves) available structural change tests assume parametric form for each curve and rely on performing independent tests of curves' parameters equality. In this paper we propose a novel nonparametric and robust test for a structural change in economic system detection namely local extension of Wilcoxon test for two functional samples. The test can be effectively used for detecting a structural change in functional time series. The underlying idea is to compare populations at different locality level, which may be interpreted as data resolution.
In our proposal the local Wilcoxon test statistic is induced by the corrected modified band depth \citep{LopezRomo} with a concept of locality proposed by \citep{Pand}.
\\  The rest of the paper is organized as follows.
In the second section we briefly sketch basic concepts of a two sample test for homogenity in the context of functional time series. Next we introduce two sample local Wilcoxon test statistic for detecting a structural change in functional time series. We discuss in Section 4 properties of the procedure via numerical simulations and test the applicability of the proposed methodology on empirical examples (i.e. internet users activity and yield curves monitoring). In the fifth section we conduct a short sensitivity analysis. The last Section 6 contains a brief summary.
\section{A concept of homogenous functional data}
\label{S:1}

An intensive debate on adequate tests of homogenity for functional data is carried in literature nowadays \citet{Flores} (see the paper and references therein). In \citet{Flores} selected two-sample homogenity tests basing on maximal depth elements comparison were discussed.
\\ Functional time series are usually defined in terms of functional stochastic processes with values in Banach or Hilbert spaces \citep{Bosq,HK}. 
In \citep{Bosq} it is explained that, probability distribution $\textbf{F}$ of functional random variable does exist. We look at the random curve $X=\{X(t), t\in [0,T]\}$ as a random element of the space $L^2=L^2([0,T])$ equipped with the Borel $\sigma-$algebra.
The $L^2$ is a separable Hilbert space with the inner product $<x,y>=\int x(t)y(t)dt.$ 
\newline We consider a sample of curves but each curve is observed at discrete and finite grid of points in practice. Discrete data are transformed into curves using various techniques including i.e. nonparametric smoothing (see \citet{Ra}). 
\newline In their book \citet{HK} prove a lot of properties of the functional estimators and among all they show that, under some regularity conditions, mean value and variance are unbiased and mean square error consistent estimators.

\citet{HKR} formalize the assumption of stationarity in the context of functional time series and propose several procedures to test the null hypothesis of stationarity, which in turn may be used to detect a structural change in FTS (functional time series) setup.
Furthermore \citet{HKR} have noted that spectral analysis of nonstationary functional time series has not been developed to a point where usable extensions could be readily derived, so they developed a general methodology for testing the assumption that a modeled functional time series is indeed stationary and analyzed the behavior of the tests under several alternatives, i.e. change point alternative.
The tests developed by the authors are consistent against any other sufficiently large departures from stationarity and weak dependence.
They warn that in the functional setting, there is a fundamentally new aspect, i.e. convergence of a scalar estimator of the long run variance must be replaced by the convergence of the eigenvalues and the eigenfunctions of the long run covariance function.
They note that their method
is extremely computationally intensive.
\newline An obvious increasing/decreasing trend is doubtful to obtain in functional data. \citet{Frai} considered functional time series, where a trend is expected. They defined different kinds of trend and then show test that enable detecting them.
The authors developed the nonparametric tests for the proposed increasing trends for a sequence of functional data and established their results for a multiple time series of functional data.
\\
\\
Let pose our hypotheses.
If $\textbf{F}$ and $\textbf{G}$ denote a probability distribution of the first and second population, respectively, we can formulate null and alternative hypothesis:
\begin{equation}
H_0: \textbf{F}=\textbf{G}
\textit{    vs.    }
H_1: \textbf{F}\neq \textbf{G}.
\end{equation}
Our first aim is to test the null hypothesis against its alternative having two samples in a disposal. In this situation our first set of hypotheses states that two samples are drawn from the same distribution, while the alternative states the opposite.
We use a local Wilcoxon statistic to deal with the problem. 
\newline Our second aim is to use the local Wilcoxon statistic to detect a structural change in functional time series. In other words we intend to test a set of the following hypotheses:
\begin{equation}
H_0: \textbf{F}_{X_1}=\textbf{F}_{X_2}=...=\textbf{F}_{X_N}
\textit{ vs.}
H_1: \textbf{F}_{X_1}=...=\textbf{F}_{X_k}\neq \textbf{F}_{X_{k+1}}=...=\textbf{F}_{X_N}
\end{equation}
for some $k \in\{1,2,...,N\}$,
where $\textbf{F}_{X_i}$ is a probability distribution of a functional random variable $X_i$. 
\newline We use a moving local Wilcoxon statistic for the purpose.

\section{Our proposals}
In the following section we introduce a two sample local Wilcoxon test for homogenity i.e. for veryfying (1) set of hypotheses. 
\subsection{Ranks Induced by Depth Functions}
Consider a FDA setup in which each observation is a real function %${{x}_{i}}$ $i=1,...,n$ 
defined on a common interval in $\mathbb{R}$.
In order to introduce rank based statistic for comparing samples of functional data, we focus our attention on statistical depth functions for functional objects. It enables us for ordering these objects in terms of departure of an object from a center - the functional median. 
\emph{The data depth concept} was originally introduced as a way to generalize the concept of order statistics to a multivariate case (see \citet{Mos}), but presently is treated as a very powerful data analytic tool which is able to express various features of the underlying distribution. The depth function yields information about spread, shape, and asymmetry of a distribution, through depth regions (see: \citet{LPS, Mos} and references therein).
Within the depth concept it is possible to propose effective methods of location and scale differences testing (see \citet{Liu}). \\
Classical depth functions associate with any center of symmetry a maximal depth value. Together with the fact that depth decreases along any halfline originating from any deepest point, this leads to nested star-shaped (in most cases convex) depth regions, whatever the underlying distribution may be nonconvex (\citet{ZS}).
Distributions that are multimodal or have nonconvex support however are present in many economic applications (mixture models, multi-regime time series or issues solved by means of clustering procedures). These facts motivated several authors to extend the concept of depth to make it flexible enough to deal with such distributions. Such extensions are available in the literature, under the name \emph{local depths}. In this paper we use the concept of local depth proposed by \citet{Pand} and implemented among others in \citet{KZ}.
Thorough presentation of the depth concept may be found in \citep{ZS,Mos,NR}.\\
In recent years, some definitions of depth for functional data have been proposed as  well. \citet{Frai1} considered a concept of depth based on the integral of univariate depths, \citet{Lopez} introduced functional depths taking into consideration a shape of considered curve. A very useful theoretical considerations related to a definition of the functional depth and comparative study of several functional depths may be found in \citet{NR}.
\begin{figure}
\centering
\begin{minipage}[t]{.45\textwidth}
\centering
\includegraphics[width=.95\linewidth]{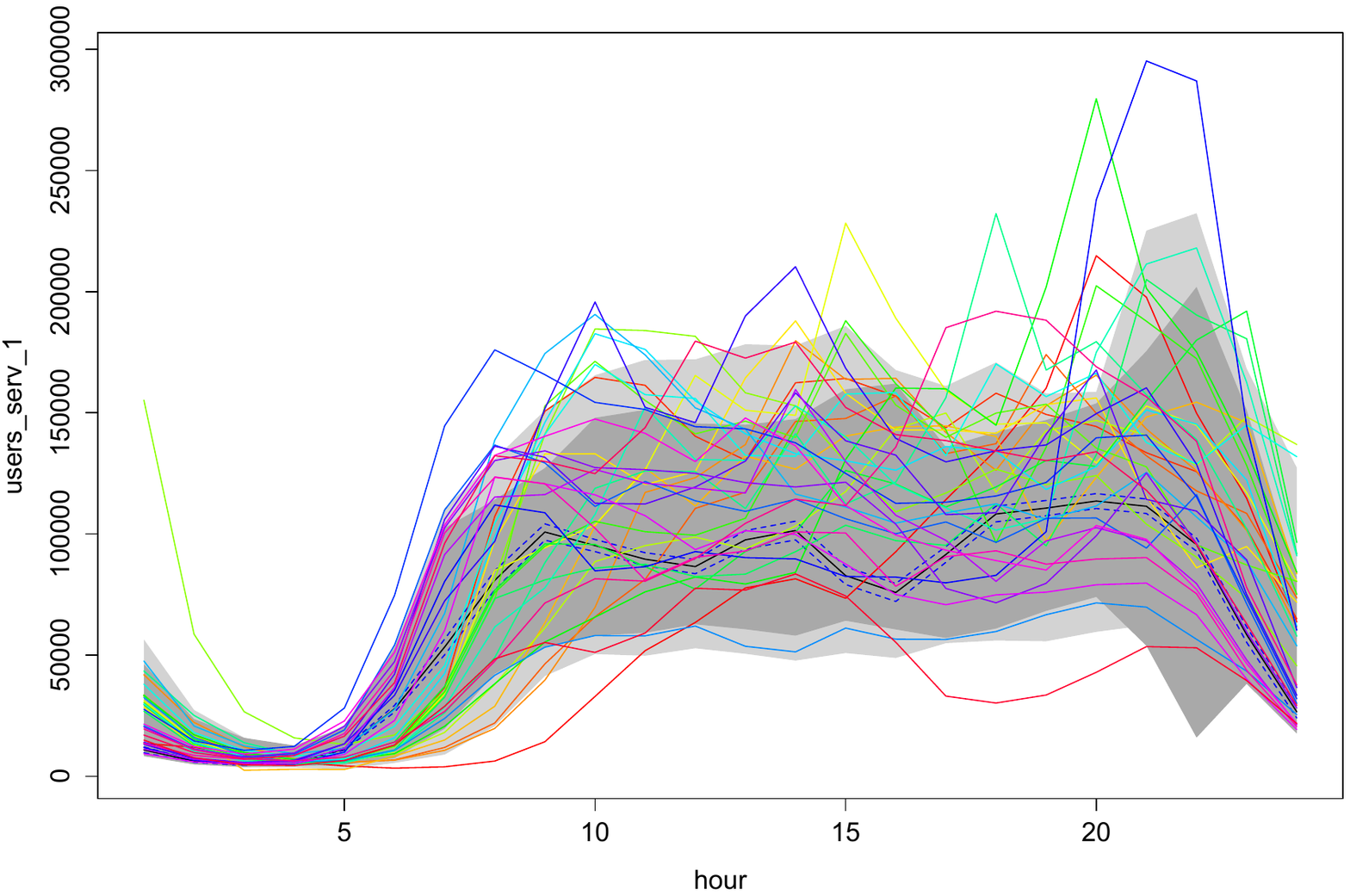}
\caption{Functional boxplot for numbers of users in service 1 during day and night.}
\label{fig1}
\end{minipage}
\mbox{\hspace{0.1cm}}
\begin{minipage}[t]{.45\textwidth}
\centering
\includegraphics[width=.95\linewidth]{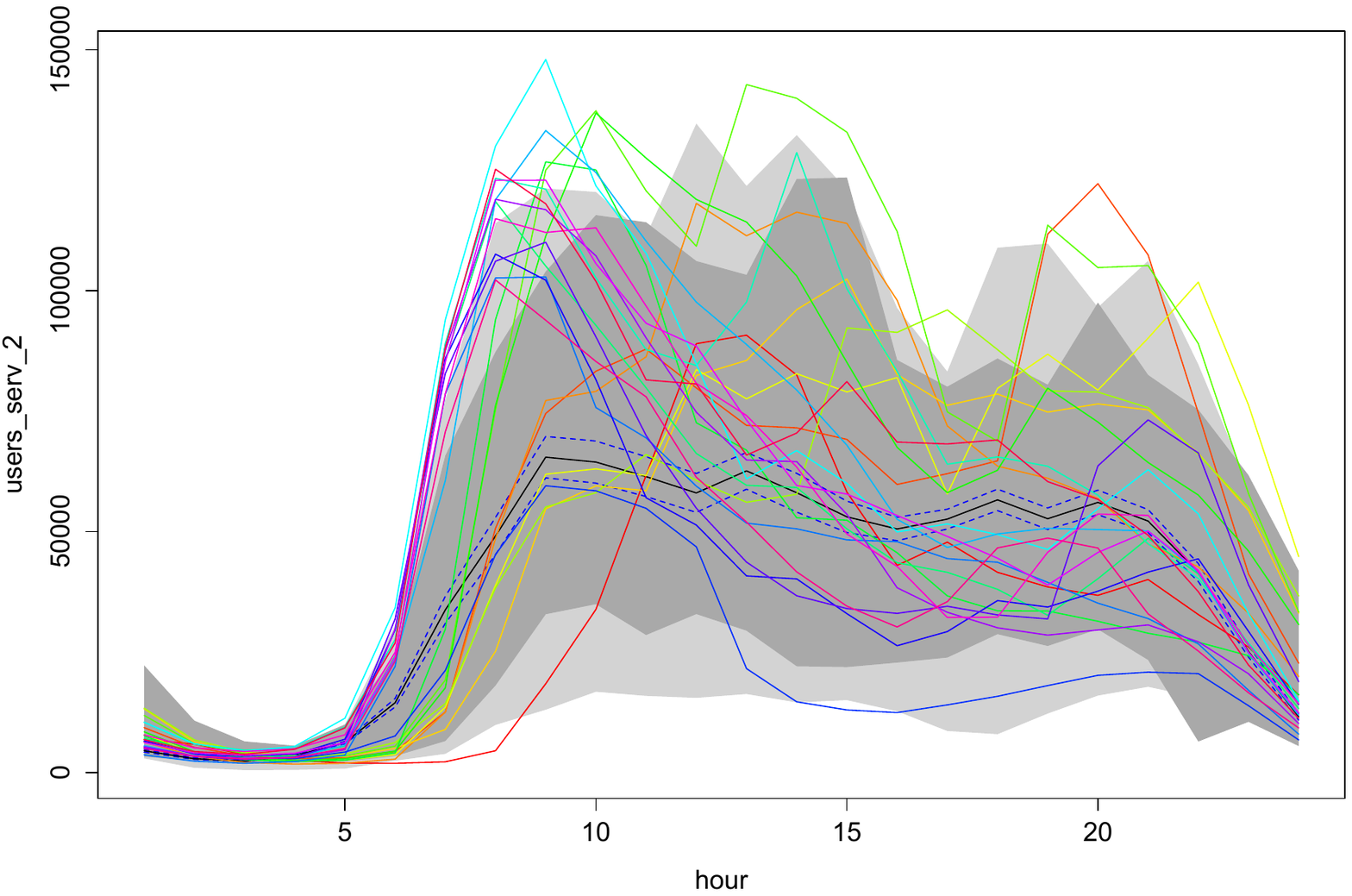}
\caption{Functional boxplot for numbers of users in service 2 during day and night.}
\label{fig2}
\end{minipage}
\end{figure}
In our opinion statistics induced by the functional depths may effectively be used for non-parametric and robust monitoring of certain properties of functional time series.
In this context we propose to use a novel tools offered by robust functional analysis to test a reasonable hypothesis of equality distributions of the two given sets of functional sequences. Consider a situation in which we would like to compare two functional sequences $\{X_i\}_{i=1}^{n-m}$ and $\{Y_i\}_{i=1}^m$.
\newline We suggest to proceed in a following manner. Using a concept of corrected generalized band depth (see \citep{Lopez}), we rank the original observations from the observation which is the closest to the functional median up to the observation which is the furthest one. The Wilcoxon test or another rank test is conducted then (see \citep{Hajek} for alternative rank tests)
\newline Let examine our procedure in details. Firstly, we combine both samples $\{X_i\}_{i=1}^{n-m}$ and $\{Y_i\}_{i=1}^m$. Let now $\mathbf{X}=\{x_1,...,x_n\}$ denote a combined sample of continuous curves defined on the compact interval $T$. Let $\lambda$ denote the Lebesgue measure and let $a(i_1,i_2)=\{t\in T : x_{i_2}-x_{i_1}\geq 0\}$, where $x_{i_1}$ and $x_{i_2}$ are band delimiting objects. Let $L_{i_1,i_2}=\frac{\lambda(a(i_1,i_2))}{\lambda(T)}$.
 \emph{A corrected generalized band depth} of a curve $x$ with respect to the sample $\mathbf{X}$ is (see \citep{Lopez,LopezRomo}])
\begin{equation}
cGBD(x|\mathbf{X})=\frac {2}{n(n-1)}\sum_{1\leq i_1<i_2\leq n}\frac{\lambda(A^c(x;x_{i_1},x_{i_2}))}{\lambda(T)}
\end{equation}
where
$$A^c(x;x_{i_1},x_{i_2})=
\{t\in a(i_1,i_2) : x_{i_1}(t)\leq x(t)\leq x_{i_2}(t)\},  \textrm{ if } L_{i_1,i_2}\geq \frac 12 $$
and
$$A^c(x;x_{i_1},x_{i_2})=\{t\in a(i_2,i_1) : x_{i_2}(t)\leq x(t)\leq x_{i_1}(t)\},  \textrm{ if } L_{i_2,i_1}> \frac 12.
$$
Band depth is thus modified so as to consider only the proportion of the domain where the delimiting curves define a contiguous region which has non-zero width.
To conduct the construction we evaluate the depth regions of order $\alpha$ for cGBD, i.e.
$$R_{\alpha}(P)=\{x : cGBD(x,P)\geq\alpha\}.$$
For any depth function $D(x,P)$, the depth regions, ${{R}_{\alpha }}(P)=\{x\in L^{2}([0,T]):D(x,P)\geq \alpha \}$ are of paramount importance as they reveal very diverse characteristic of probability distribution $P:$ location, scatter, dependency structure (clearly these regions are nested and inner regions contain larger depth).
When defining local depth, following the concept of \citet{Pand}, it will be more appropriate to index the family $\{{{R}_{\alpha }}(P)\}$ by means of probability contents.
Consequently, for any $\beta \in (0,1]$ we define the smallest depth region with P-probability equal or larger than $\beta$ as
$$R^{\beta}(P)=\bigcap_{\alpha\in A(\beta)}R_{\alpha}(P), $$
where $A(\beta)=\{\alpha\geq0 : P(R_{\alpha}(P))\geq\beta\}$.
The depth regions ${{R}_{\alpha }}(P)$ or ${{R}^{\beta }}(P)$ provide neighborhood of the deepest point only.
However we can replace $P$ by its symmetrized version ${{P}_x}=\frac{1}{2}{{P}^{\mathbf{X}}}+\frac{1}{2}{{P}^{2x-\mathbf{X}}}$.
We shall set a definition. Let $D(\cdot ,P)$ be a depth function. The corresponding \emph{sample local depth function at the locality level $\beta \in (0,1]$} is $L{{D}^{\beta }}(x,{{P}^{(n)}})=D(x,{{P_x}^{\beta (n)}})$, where $P_{x}^{\beta (n)}$ denotes the empirical measure with those data points that belong to $R_{x}^{\beta }({{P}^{(n)}})$. $R_{x}^{\beta }({{P}^{(n)}})$ is the smallest sample depth region that contains at least a proportion $\beta $ of the $2n$ random functions ${{x}_{1}},...,{{x}_{n}},2x-x_{1},...,2x-x_{n}$. Depth is always well defined -- it is an affine invariant from original depth. For $\beta =1$ we obtain global depth, while for $\beta \simeq 0$ we obtain extreme localization.
As in the population case, our sample local depth will require considering, for any $x\in {{\mathbb{L}}^{2}}$, the symmetrized distribution $P_{x}^{n}$ which is empirical distribution associated with ${x_{1}},\ldots,{x_{n}},2x-{x_{1}},\ldots,2x-{x_{n}}$.
Sample properties of the local versions of depths result from general findings presented in \citep{ZS}.
\newline Implementations of local versions of several depths including projection depth, Student, simplicial, $L^p$ depth, regression depth and modified band depth can be found in free R package \emph{DepthProc} (see \citet{KZ}). For choosing the locality parameter $\beta$ we recommend using cross validation related to an optimization a certain merit criterion (the resolution being appropriate for comparing phenomena in terms of their aggregated local shape differences, that relies on our knowledge on the considered phenomena).
\subsection{Local Wilcoxon Test for testing homogeneity}
Let us consider two samples and  $\mathbf{X}=\{{x_1}, {x_2},...,{x_n} \}=\{X_i\}_{i=1}^{n-m}\cup\{Y_i\}_{i=1}^m$.
The ranks induced by a local corrected generalized band depth with prefixed locality parameter $\beta \in (0,1]$ are
\begin{equation}
R_l=\#\left\{ {x_{j}}\in \mathbf{X}:cGDB(\beta )({{x}_{j}},\mathbf{X})\le cGDB(\beta )({{x}_{l}},\mathbf{X}) \right\},
\end{equation}
$l=1,...,n.$
Ranking the original observations according to the $cGBD$ is done subsequently. Let the unified ranks in the combined sample of all observations be $R_l$, $l=1,2,...,n$ or $R_{x_1},...,R_{x_{n-m}}$ ranks of $X_i$'s and $R_{y_1},...,R_{y_m}$ ranks of $Y_i$'s.
\\
\textbf{PROPOSAL 1.} We propose to conduct a proper Wilcoxon test to test a hypothesis of equality of the two distributions generating two given sets of functional sequences. The $\beta$--local (two independent samples) Wilcoxon rang sum statistic for functional data takes a form
\begin{equation}
S^{\beta}=\sum\limits_{i=1}^{n-m}{{{R}_{x_i}}},
\end{equation}
where ranks are induced by local cGBD with locality parameter $\beta$.
Following Li and Liu (2004) it is worth noticing  that having in a disposal two samples X and Y and any depth function, one can calculate depth in the combined sample $X\cup Y$, assuming empirical distribution calculated using all observations, or calculating this distribution assuming only one of the samples X or Y. If we observe that depths for ${X}_{l}'s$ indicate the center of the combined sample and depths for ${{Y}_{l}}'s$ indicate peripheries we conclude Y was taken from distribution with bigger scatter. Generally speaking, differences in allocations of ranks between samples indicate various differences in shapes of underlying distributions and hence a departure from their equality \citep{LPS}.
In the functional setting a difference in scale means that $\alpha-$ central region drawn from the population X consists of smaller amount of probability mass than taken from the population Y and hence Y is more scattered than X.
The locality parameter $\beta$ indicates a resolution in which we compare the populations. It ranges from a very misty comparison (parameter close to one) to a very sharp comparison (the parameter close to zero).
From other point of view we can treat the statistic (5) as an aggregate representing local asymmetry in data set (see \citep{Pand}). Differences in value of (5) for two samples indicate differences in aggregated local asymmetry but simultaneously in local location and scale.
Notice, that for $\beta=1$ we have to do with classical Wilcoxon rank sum test and hence we can use tables for this test to obtain critical values, and use well known ties breaking schemes in case of ties (see \citet{Jure}). Big or small values of test statistics indicate differences in distributions between the samples then. For other $\beta$ values, for each point we calculate depth w.r.t. empirical distribution symmetrized in this point. It may happen that two points have the same depth value and hence the same rank. We expect significant differences in sums of ranks for samples drawn from different continuous distributions however (different distributions should be characterized by different kinds of local asymmetry).  The differences we underline are related to the parameter of resolution $\beta$ in which we conduct the comparison. The $\beta$ parameter on the other hand may be treated as parameter of data peeling of the combined sample -- a parameter of desired sensitivity to contamination of our procedure.
"A power" of the test depends on differences between location and scale of underlying distributions but also on differences in "shape" of underlying distributions in appropriate functional space. For practical purposes we recommend Monte-Carlo evaluation of the "power" in case of selected alternatives being especially important for a decision maker from a merit point of view.
Merit properties of the proposal depends on properties of functional depth used.
Sample properties and other asymptotic properties of the proposed statistic result from \citep{NR} and \citep{Pand}. Notice however that \citep{NR} did not consider local functional depth but only global versions.
\\  \citet{Flores} constructed four different statistics to measure distance between two samples basing on maximal depth elements comparison. They proposed two sample tests for homogenity in the context of functional data analysis. They did not use in their considerations a concept of local depth.
Our approach enables to use a locality parameter $\beta$, which indicates a resolution in which we compare the populations. It ranges from a very misty comparison (parameter close to one) to a very sharp comparison (the parameter close to zero).
The researcher may adjust the locality parameter on the grounds of the matter being considered and her/his experience. Our proposal (5) outperforms their proposals in cases of multimodal distributions. \\ \\
\textbf{PROPOSAL 2 - detecting a structural change.} There is given a functional reference sample $X_1,...,X_M$.
We would like to compare the stream of functional data $Y_1,...,Y_N$, where $N>>M$ with our reference sample, i.e. to detect a structural change in a functional data stream.
\\ We construct a moving window of length $L$ and sequentially test a homogeneity of $X_1,...,X_N$ and $Y_k,...,Y_{k+L-1}$ for $k\in\{1,...,N-L+1\}$ using the statistic (5). Our procedure is able to detect a structural change in a functional data stream.
In order to obtain sample distribution of the test statistic and in consequence the necessary p-values we propose to use a maximal entropy bootstrap methodology proposed by \citet{Vinod} and implemented in \emph{meboot} R package. Note that in the time series setting due to the temporal dependence between observations, resampling and especially bootstrap seem to be the only solution to conduct statistical inference (see \citet{Shang2016}). Having empirical time series under our study we generate bootstrap samples using \emph{meboot} R package, and then we calculate our sample Wilcoxon statistic distribution to obtain appropriate p-values.

\section{Properties of the proposals -- simulation studies}
In order to check finite sample properties of our proposal we conducted simulation studies.
In order to check finite sample properties of our  "static" proposal (5) we conducted extensive simulation studies involving various differences in location and or scale and or shape of distributions generating samples. We 100 times generated two samples from the same distribution (situation representing null hypothesis) and from two distributions of the same kind but differing w.r.t. location and or scale. Similarly as \citet{HKR} and \citet{Dider} we considered samples with functional errors being generated by Wiener process and Brownian bridge divided into 1440 and 120 time points (24 hours divided into 1min and 12min time segments). We considered samples of equal and different sizes. Generally speaking in case of simple differences in location and scale our proposal performed comparable to proposals introduced in \citet{Flores} based on maximal depth elements in two samples comparison but significantly outperformed them in cases of existence of multimodality -- local differences between samples.  Fig. 3 presents sample of 50 curves generated from Wiener process observed at 120 points. Fig. 4 presents sample of 50 curves generated from 5\% mixture of two Wiener processes differing w.r.t.location. Fig. 5 presents estimated density of the statistic (5) under hypothesis that both samples are generated from the population related to fig. 3 and fig. 6 presents the estimated density of statistic (5) under alternative in which the first sample is generated from population related to fig. 3 and the second population is generated from a population related to fig. 4. It is easy to notice that the estimated densities differ w.r.t. the location and hence may be used to discriminate between populations. Further results and R codes are available upon request (we performed sensitivity analysis similar as in \citet{Flores}). 
\begin{figure}
\centering
\begin{minipage}[!ht]{.45\textwidth}
\centering
\includegraphics[width=.95\linewidth]{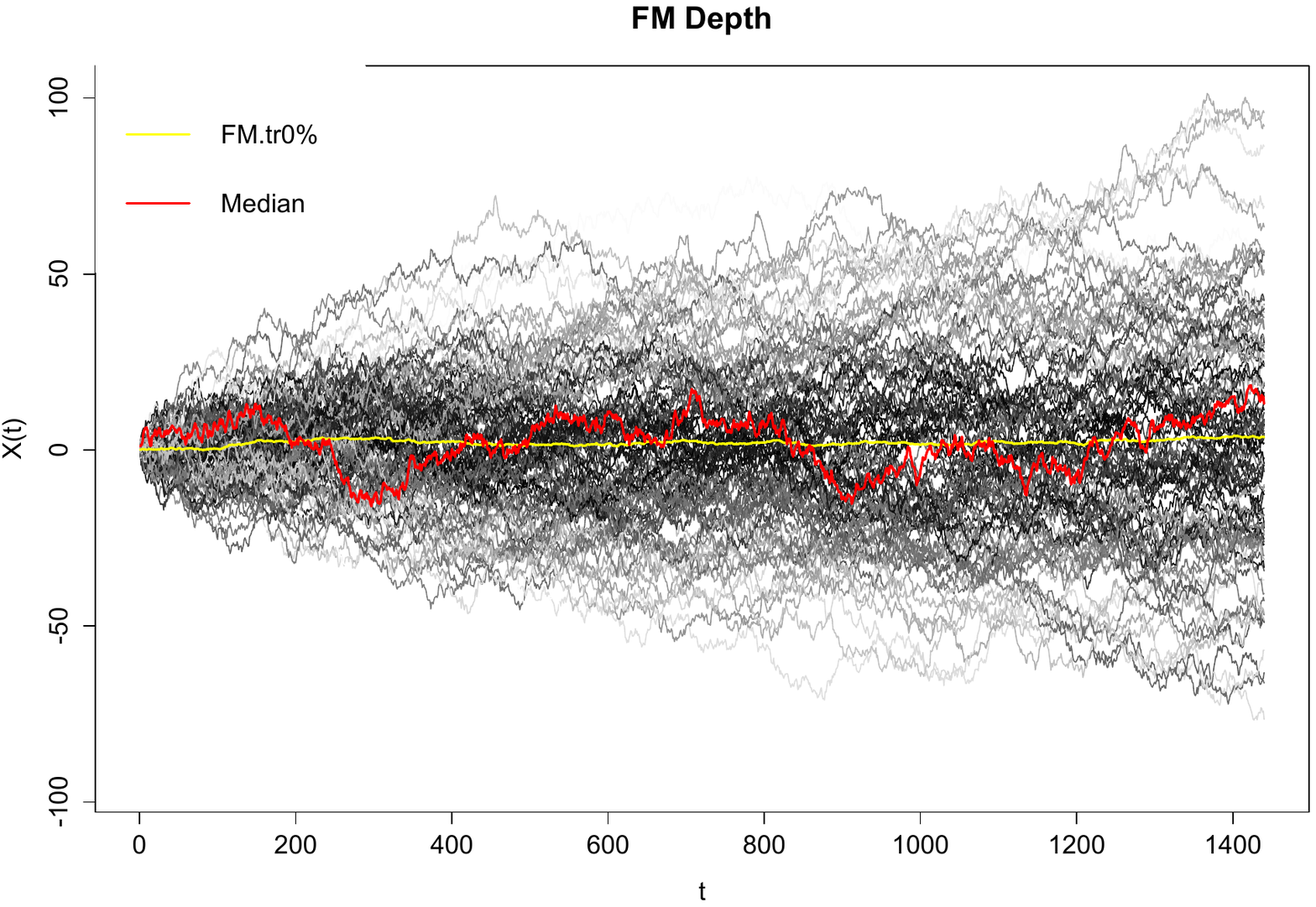}
\caption{Sample of 50 curves generated from Wiener process observed at 120 points.}
\label{fig3}
\end{minipage}
\mbox{\hspace{0.1cm}}
\begin{minipage}[!ht]{.45\textwidth}
\centering
\includegraphics[width=.95\linewidth]{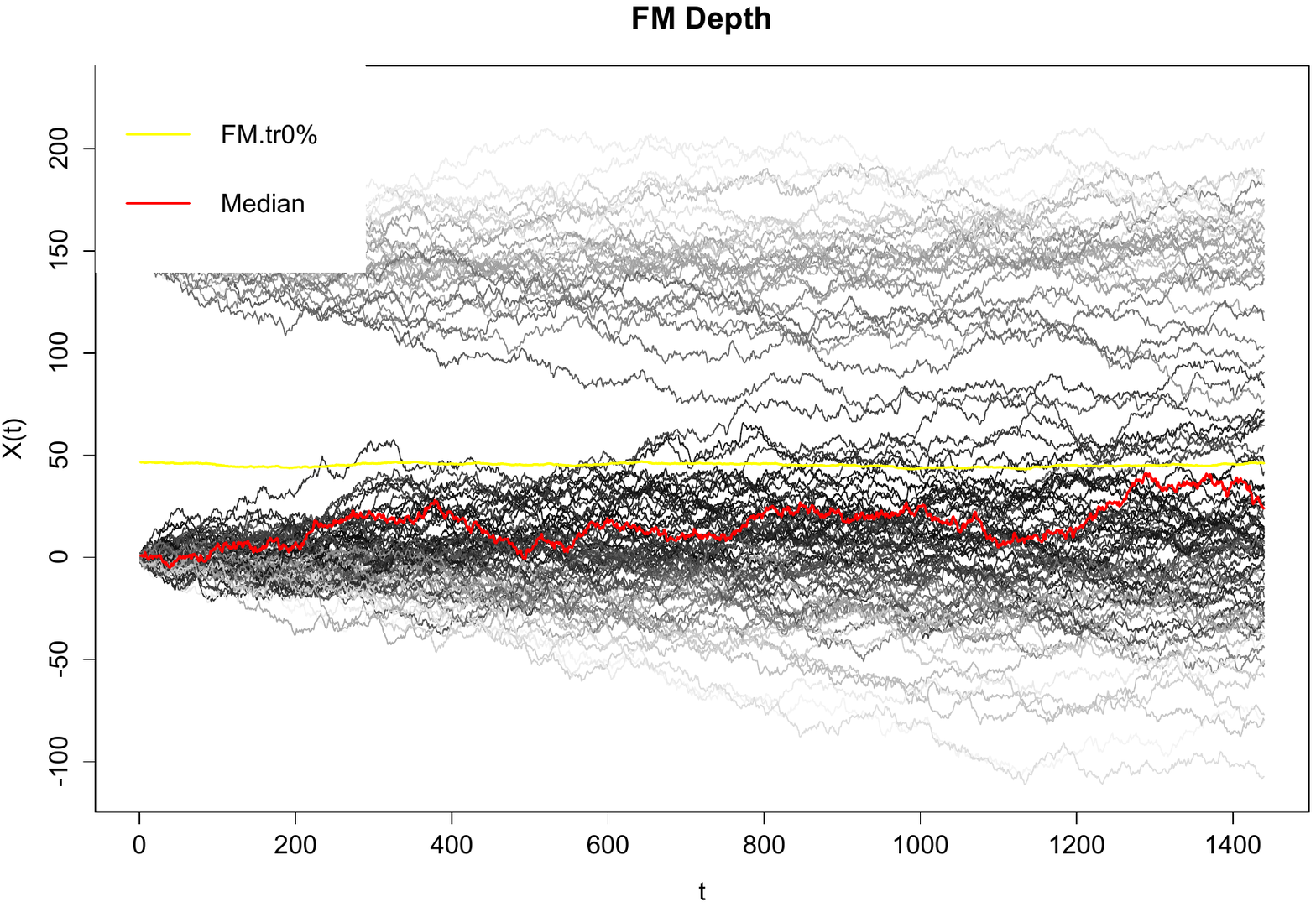}
\caption{Sample of 50 curves generated from 5\% mixture of two Wiener processes differing w.r.t.location.}
\label{fig5b}
\end{minipage}
\end{figure}

\begin{figure}
\centering
\begin{minipage}[!ht]{.45\textwidth}
\centering
\includegraphics[width=.95\linewidth]{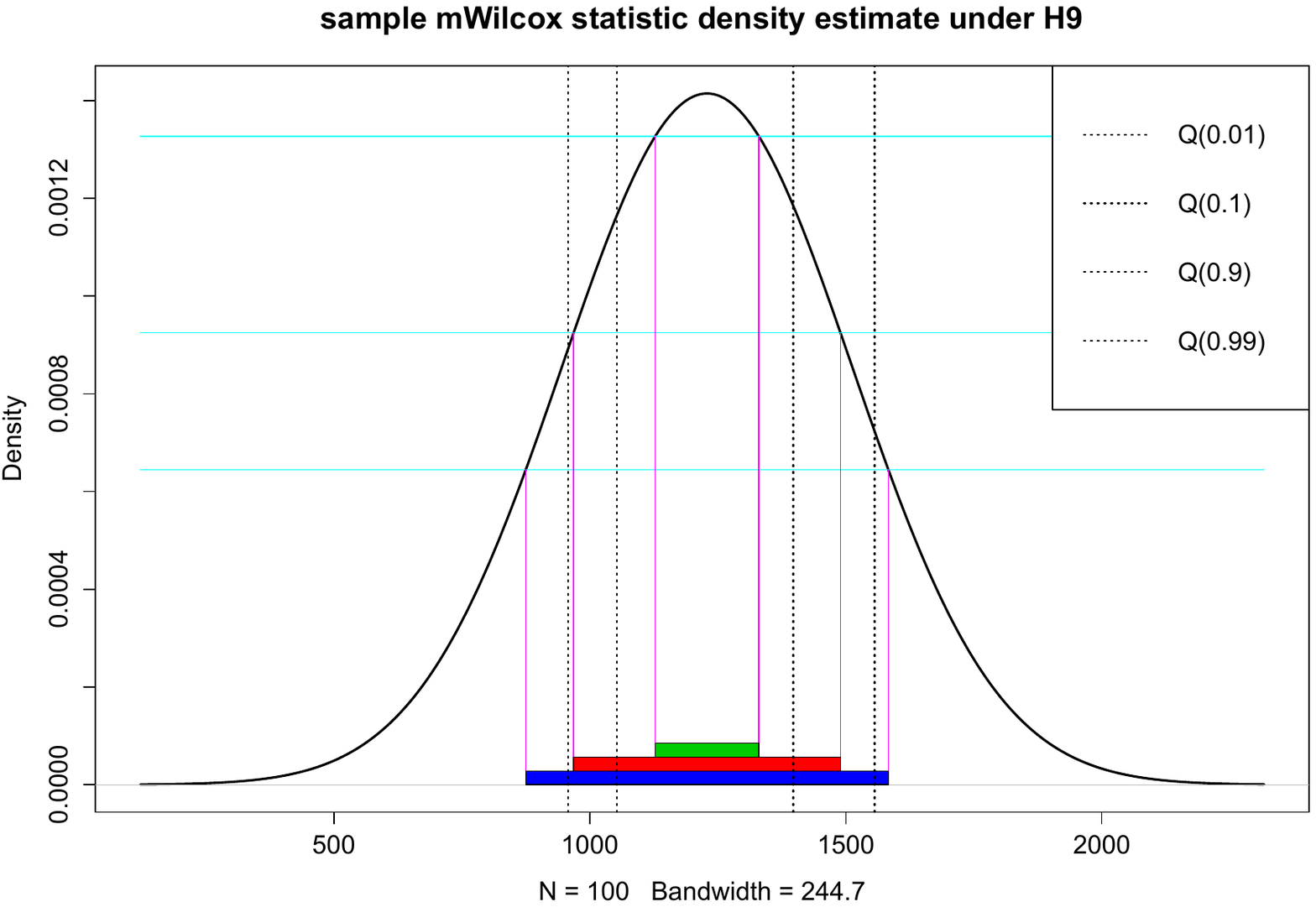}
\caption{Estimated density of the statistic (5) under hypothesis that both samples are generated from the population related to fig. 3.}
\label{fig5}
\end{minipage}
\mbox{\hspace{0.1cm}}
\begin{minipage}[!ht]{.45\textwidth}
\centering
\includegraphics[width=.95\linewidth]{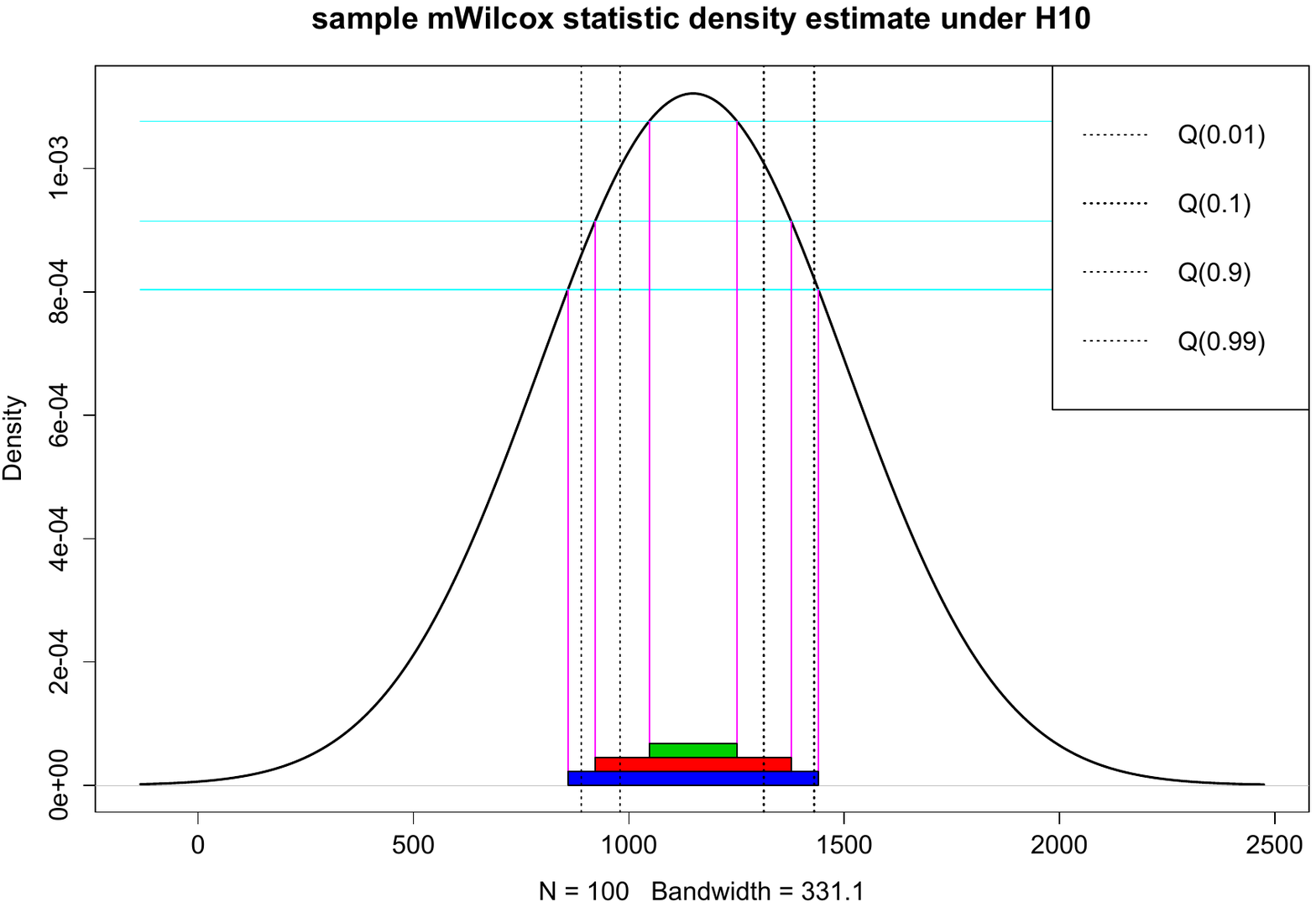}
\caption{Estimated density of statistic (5) under alternative in which the first sample is generated from population related to fig. 3 and the second is generated from a population related to fig. 4.}
\label{fig6}
\end{minipage}
\end{figure}

For checking the proposed structural change detection procedure we generated time series from the following models having economic justification in a context of cyclical properties modeling. We used functional autoregression model FAR(1), i.e. $X_{n+1}=\Psi(X_n)+\epsilon_{n+1}$, in which the errors $\epsilon_n$ and the observations $X_n$ are curves, and $\Psi$ is a linear operator transforming a curve into another curve.
The operator $\Psi$ is defined as
$\Psi(X)(t) = \int\psi(t, s)X(s)ds,$
where $\psi(t, s)$ is a bivariate kernel assumed to satisfy $||\psi|| < 1$, where
$||\Psi||^2 =\int \psi^2(t, s)dtds.$ 
The condition $||\Psi|| < 1$ ensures the existence of a stationary causal solution to FAR(1) equations. 
\\ The FAR(1) data generating processes series are thus generated according to model
$X_{n+1}(t) =\int_0^1\psi(t, s)X_n(s)ds + \epsilon_{n+1}(t),$ where n = 1, 2, . . . , N.
We used the following \citep{Dider} designs of a simulation study.
\begin{enumerate}
\item In experiment 1 we generated 100 curves using a Gaussian kernel $\psi(t, s) = C \exp\{-\frac12(t^2 + s^2)\},$ and errors of type (8) from \citet{Dider}
and then 100 curves using a kernel $\psi(t, s) = C.$ Fig. 7 presents an illustration for the experiment 1. \\ We repeated the whole experiment 100 times. Fig. 9 present results of the simulations for the experiment 1 using functional boxplot and FM depth correspondingly. 
\item In experiment 2 we generated 100 curves using a Gaussian kernel with appropriate constant $C$ and 100 curves from a mixture of two processes considered in the experiment 1 but differing with respect to parameters of the error term (8) taken from \citet{Dider}. Fig. 8 presents an illustration for the experiment 2 and
fig. 10 presents results of the simulations for the experiment 2 using functional boxplot and FM depth correspondingly. Left panel of fig. 11 shows sample density estimate ($H_d0$) for a situation in which samples are generated from process presented on the left panel of fig. 12, which is in turn a mixture of two processes. Right panel of fig. 11 presents sample density estimate ($H_d1$) where the first sample is generated from the mixture of processes presented on the left panel of fig. 12 and the second sample is generated from the mixture of processes presented on the right panel of fig. 12. The estimated densities differ significantly w.r.t. location and hence our procedure correctly detects change of type of mixture - this is a situation in which our procedure performs much better than proposals of \citet{Flores} taking into account the  deepest elements in the both samples.
\end{enumerate}
It is easy to notice that our procedure correctly detects the structural change appearing after 100th observation.
\\
Fig. 12 presents very interesting example of structural change in which our proposal (6) outperforms proposals based on statistics introduced in \citet{Flores}. Structural change relates to change of type of the mixture of processes generating curves.

\begin{figure}
\centering
\begin{minipage}[!ht]{.45\textwidth}
\centering
\includegraphics[width=.95\linewidth]{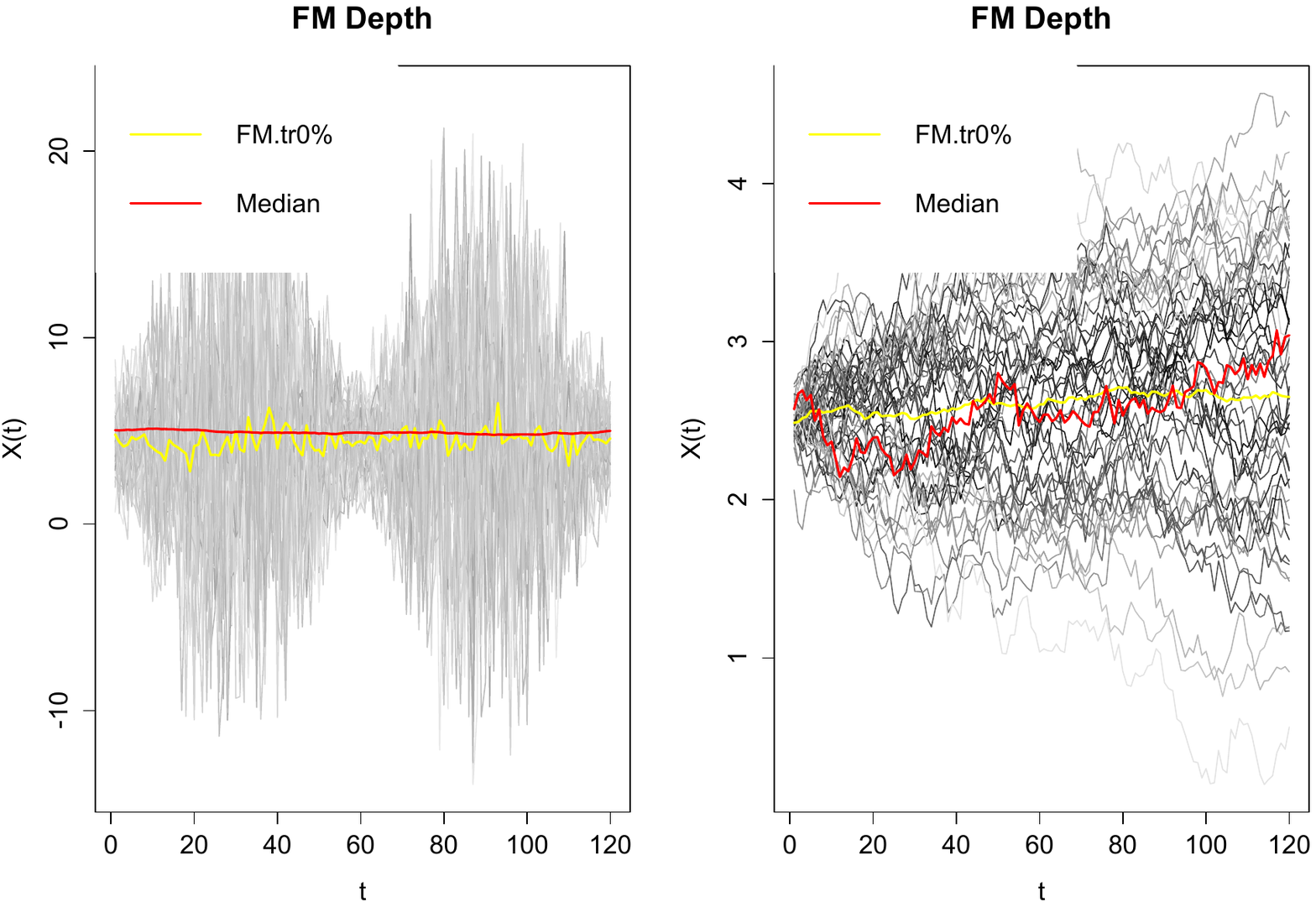}
\caption{Structural change detection in scheme 1.}
\label{fig7}
\end{minipage}
\mbox{\hspace{0.1cm}}
\begin{minipage}[!ht]{.45\textwidth}
\centering
\includegraphics[width=.95\linewidth]{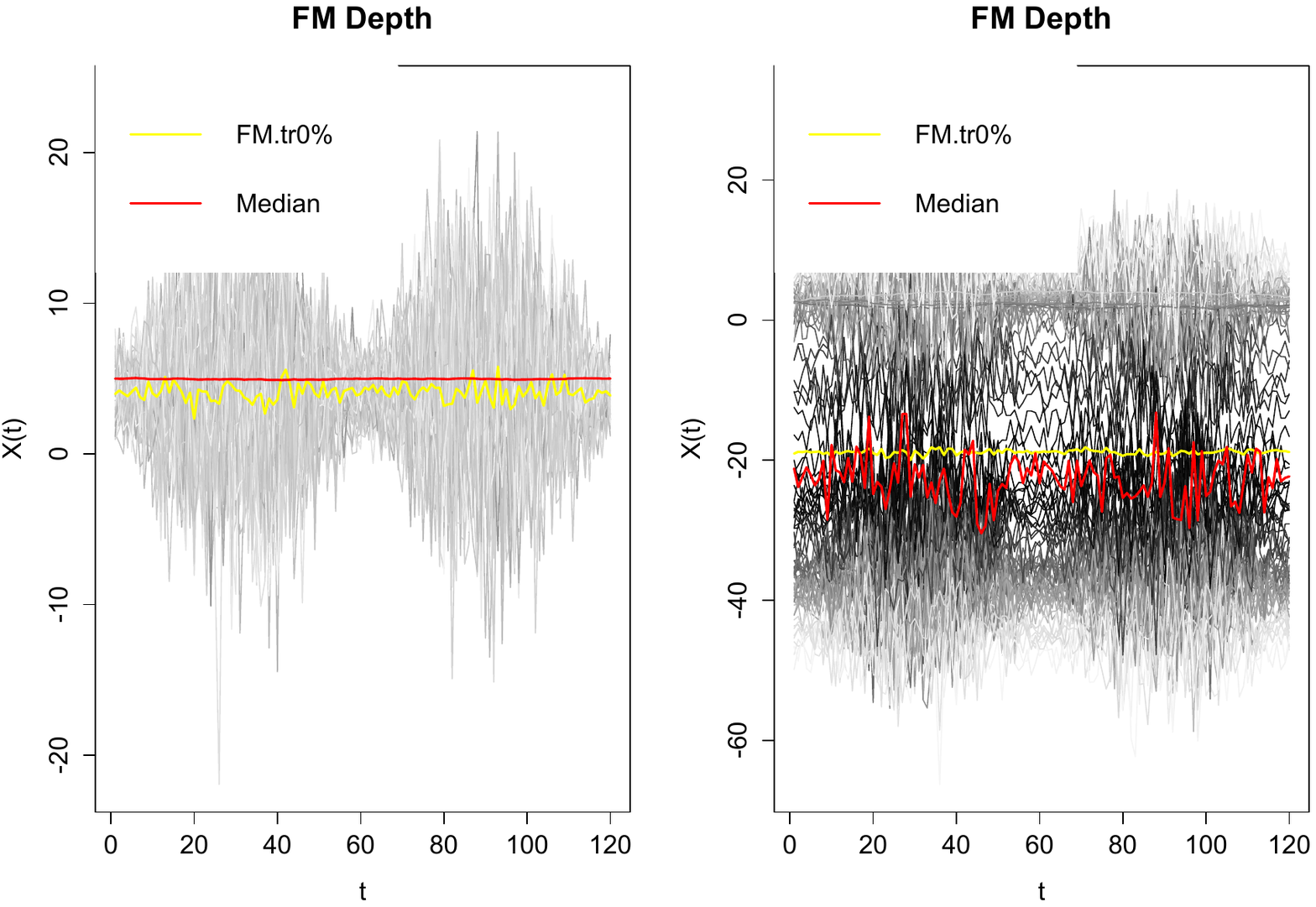}
\caption{Structural change detection in scheme 2.}
\label{fig8}
\end{minipage}
\end{figure}
\begin{figure}
\centering
\begin{minipage}[!ht]{.45\textwidth}
\centering
\includegraphics[width=.95\linewidth]{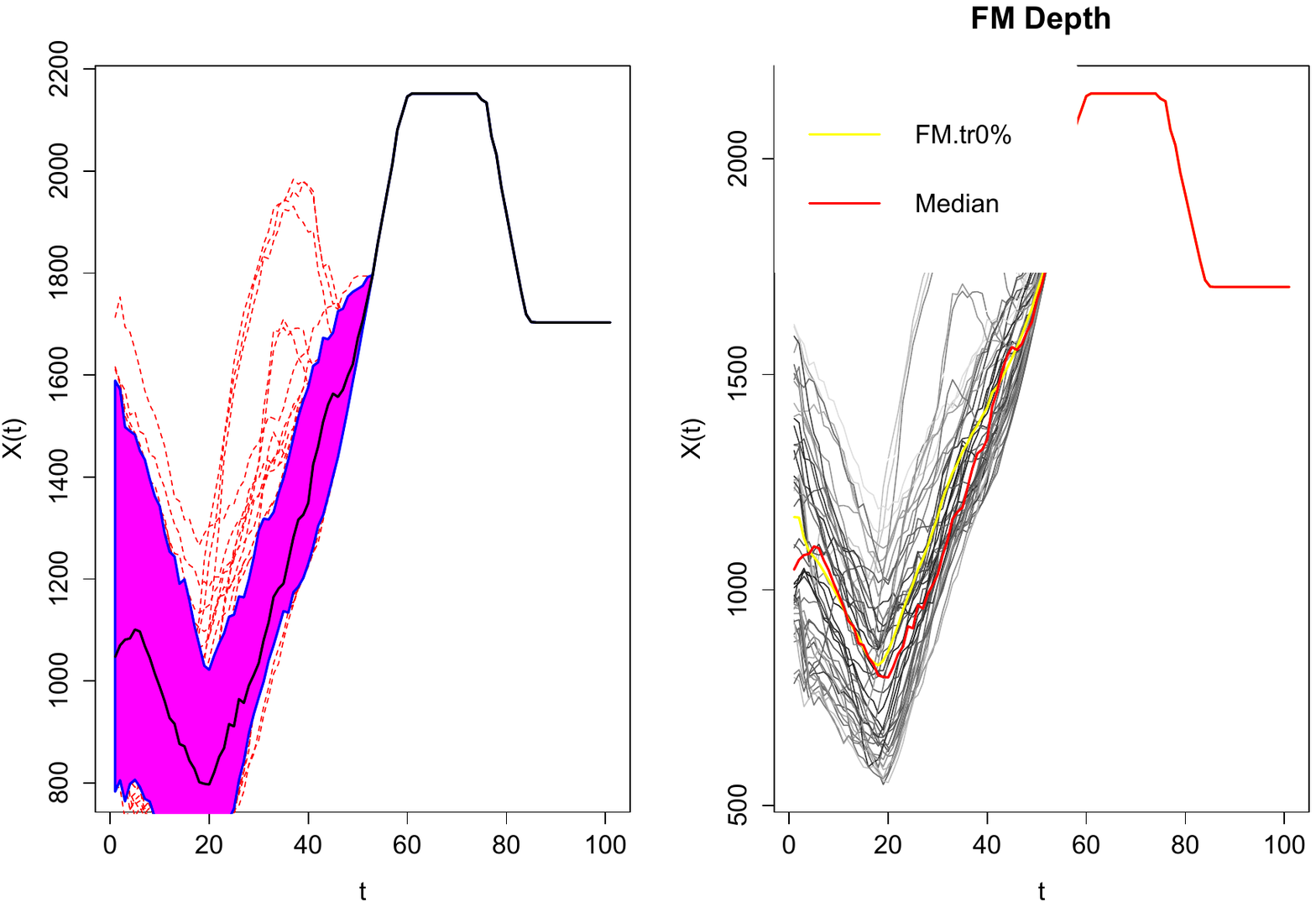}
\caption{100 detections of structural change in scheme 1 using moving Wilcoxon statistic.}
\label{fig9}
\end{minipage}
\mbox{\hspace{0.1cm}}
\begin{minipage}[!ht]{.45\textwidth}
\centering
\includegraphics[width=.95\linewidth]{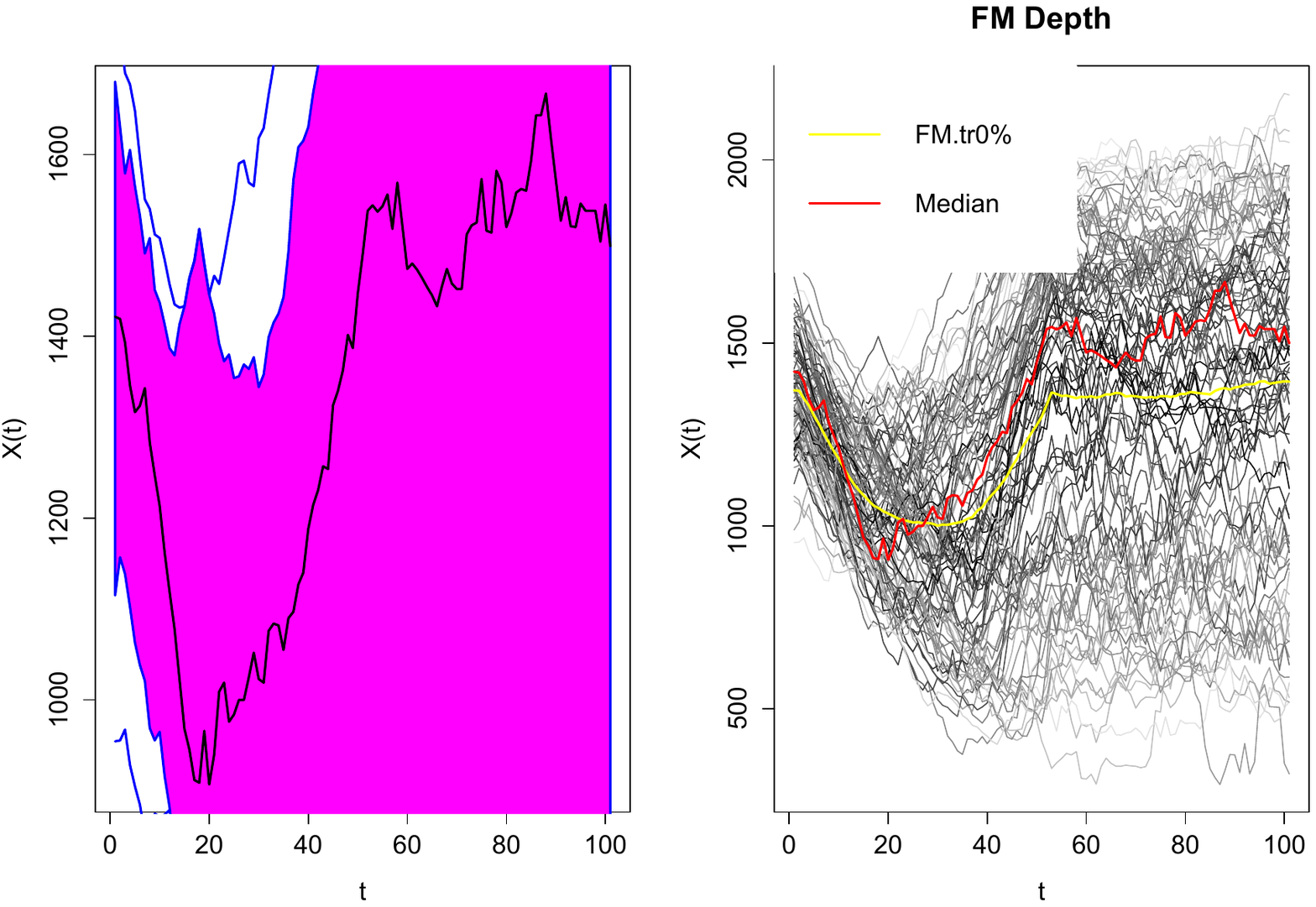}
\caption{100 detections of structural change in scheme 2 using moving Wilcoxon statistic.}
\label{fig10}
\end{minipage}
\end{figure}
\begin{figure}
\centering
\begin{minipage}[!ht]{.45\textwidth}
\centering
\includegraphics[width=.95\linewidth]{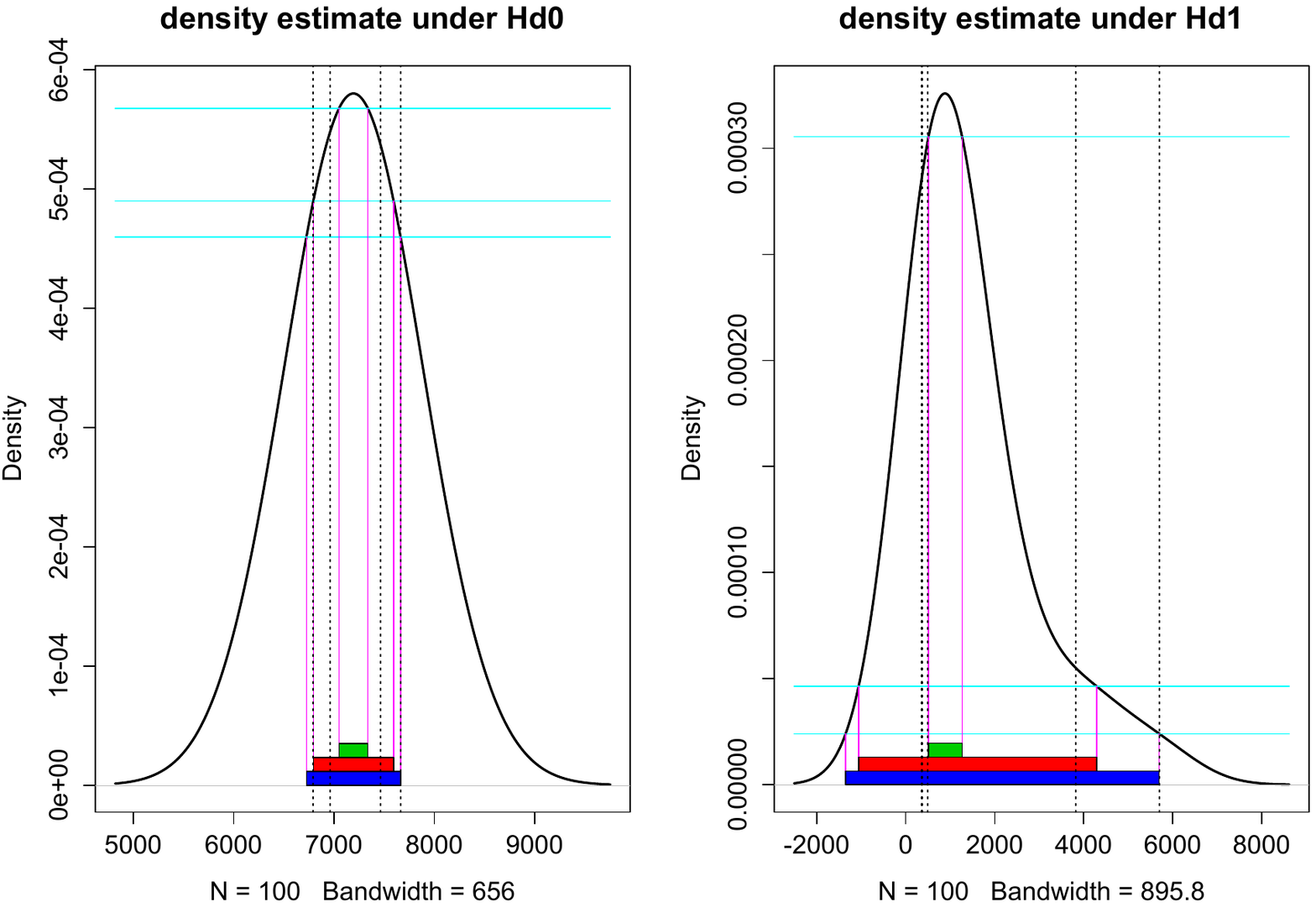}
\caption{
Sample statistic density estimate under $H_{d0}$ and under $H_{d1}$.}
\label{fig11}
\end{minipage}
\mbox{\hspace{0.1cm}}
\begin{minipage}[!ht]{.45\textwidth}
\centering
\includegraphics[width=.95\linewidth]{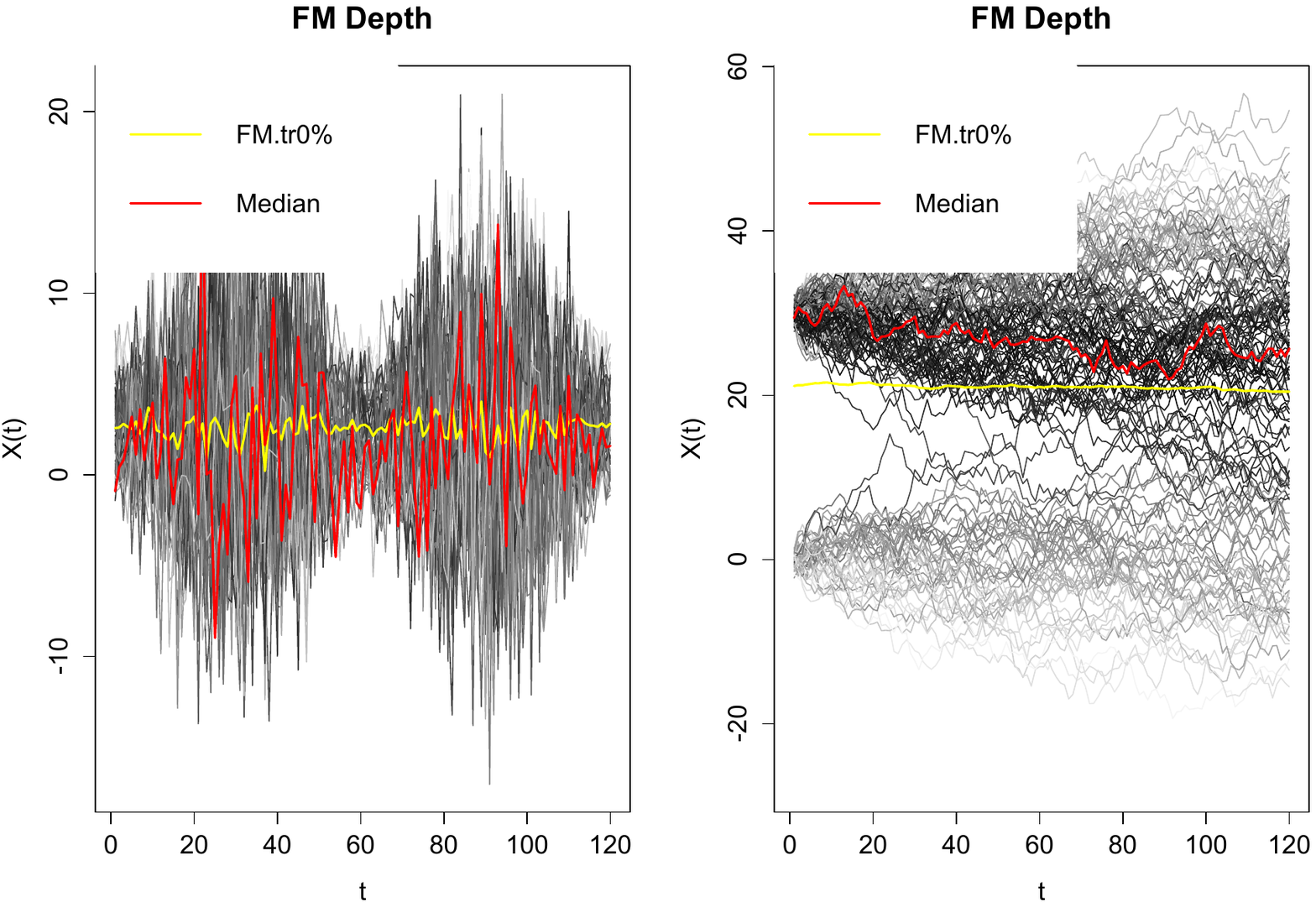}
\caption{Two processes which are mixtures of two different processes.}
\label{fig12}
\end{minipage}
\end{figure}

 \subsection{Properties of the proposal-- empirical example 1}
  For verifying empirical usefulness of the proposal we considered two Internet services with respect to number users and numbers of page views basing on real data which were kindly made for us available by owners of the services.
Fig. 1 presents functional boxplot for hourly numbers of users of the service 1 in 2013, Fig. 2 presents functional boxplot for hourly numbers of users of the service 2 in 2013.

\begin{figure}[!ht]
\centering
\begin{minipage}[!ht]{.45\textwidth}
\centering
\includegraphics[width=.95\linewidth]{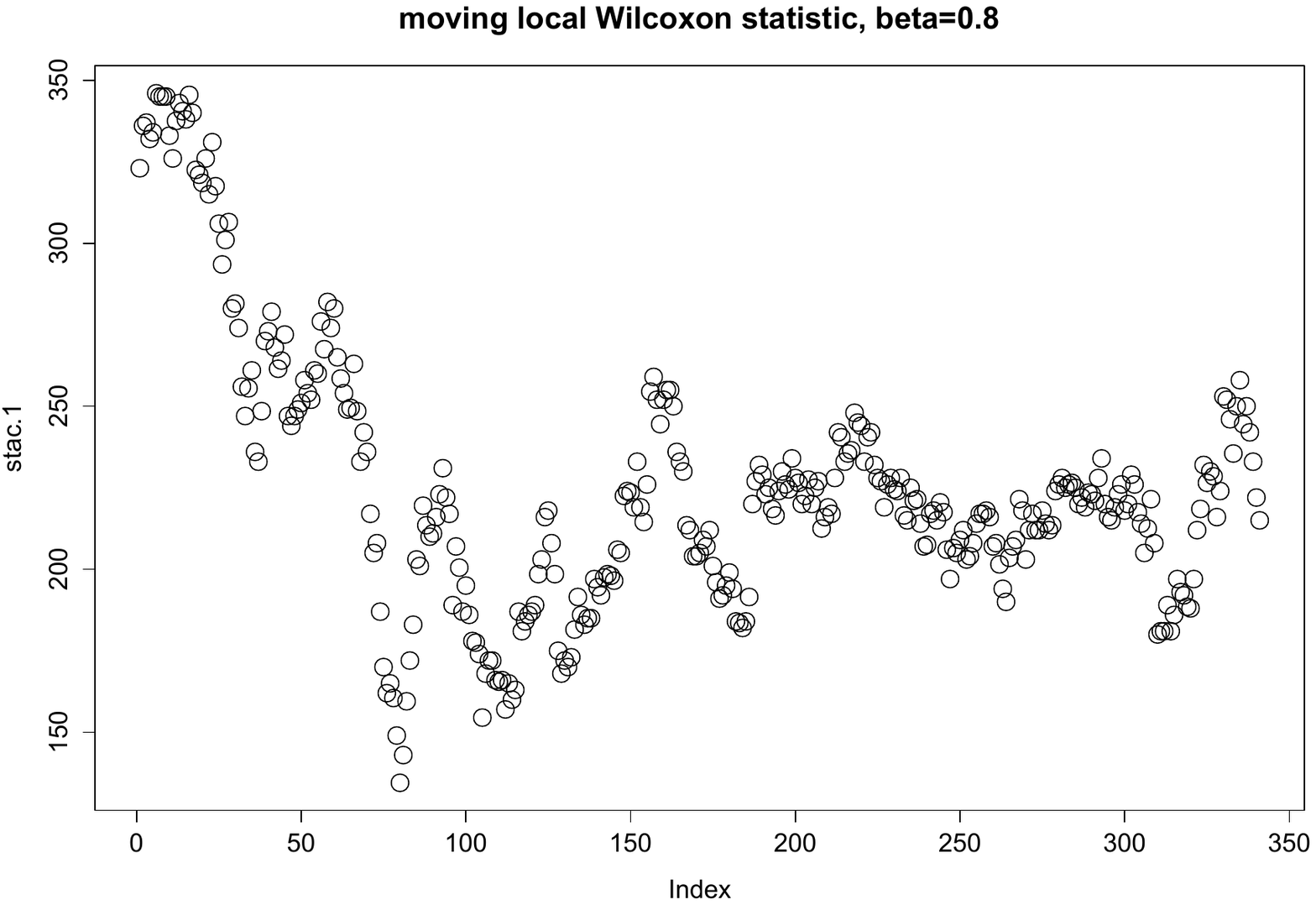}
\caption{Moving Wilcoxon statistic for numbers of users in service 1, $\beta=0.8$}
\label{fig13}
\end{minipage}
\mbox{\hspace{0.1cm}}
\begin{minipage}[!ht]{.45\textwidth}
\centering
\includegraphics[width=.95\linewidth]{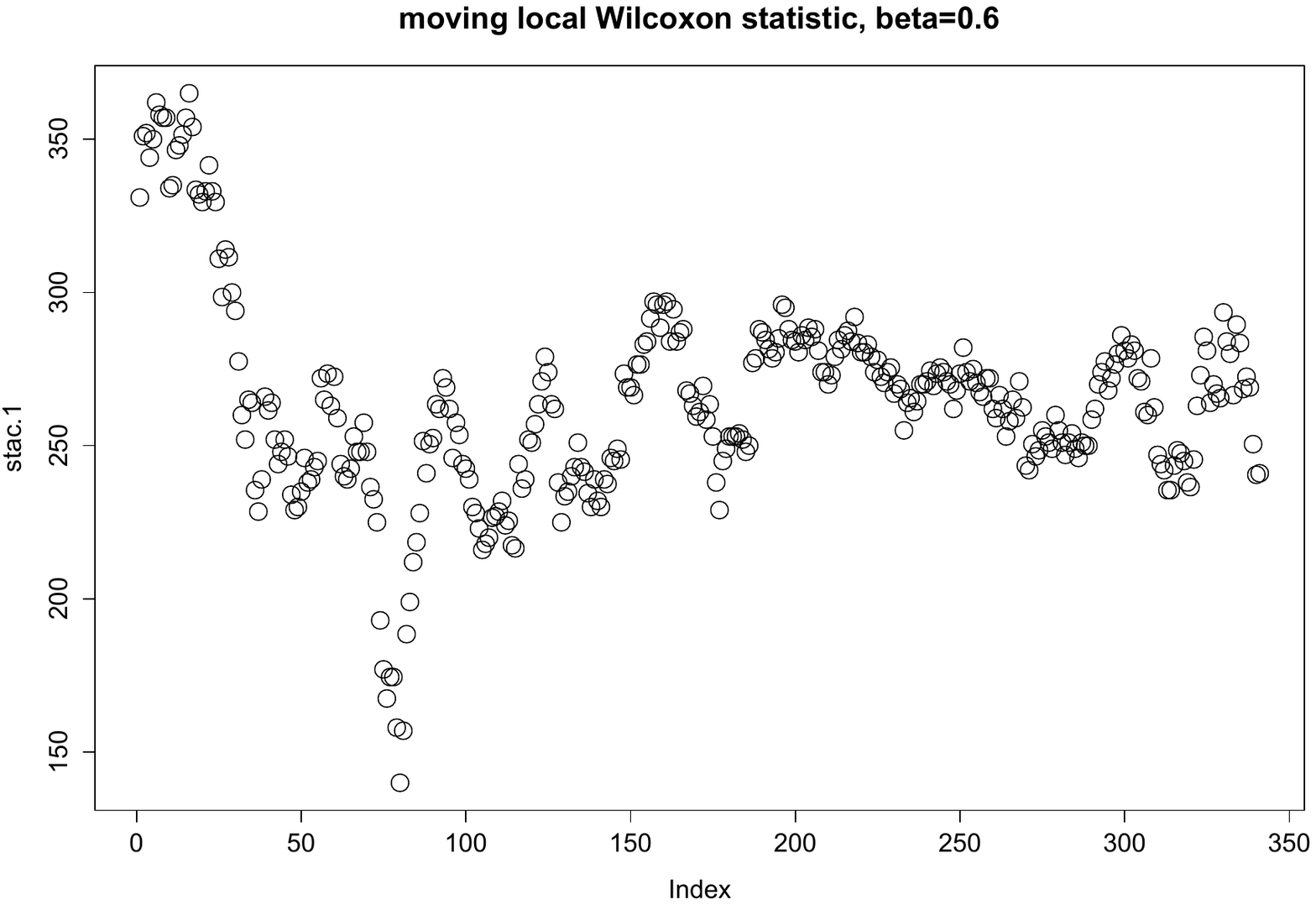}
\caption{Moving Wilcoxon statistic for numbers of users in service 1, $\beta=0.6$}
\label{fig14}
\end{minipage}
\end{figure}

\begin{figure}[!ht]
 \centering
\begin{minipage}[!ht]{.45\textwidth}
\centering
\includegraphics[width=.95\linewidth]{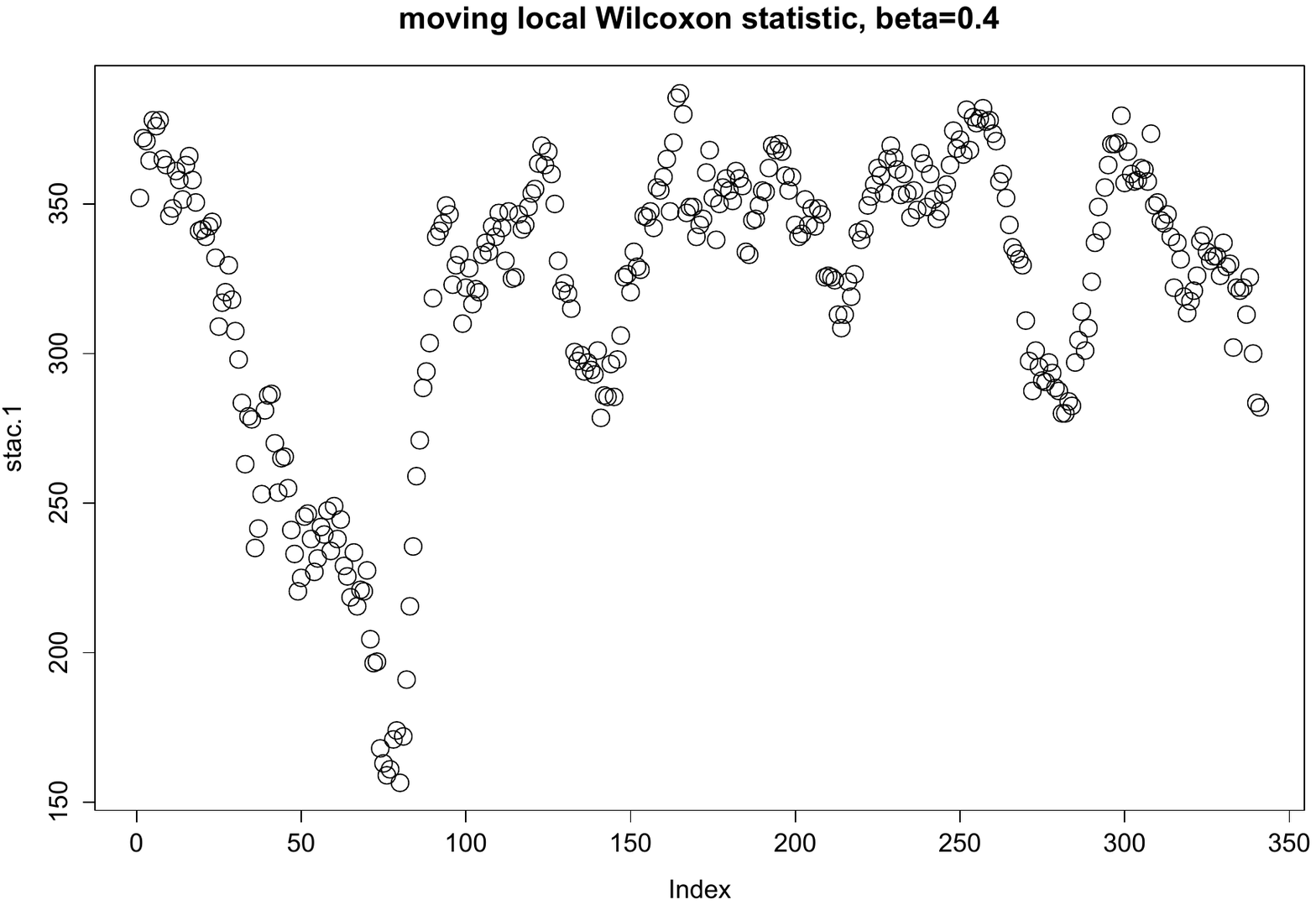}
\caption{Moving Wilcoxon statistic for numbers of users in service 1, $\beta=0.4$}
\label{fig15}
\end{minipage}
\mbox{\hspace{0.1cm}}
\begin{minipage}[!ht]{.45\textwidth}
\centering
\includegraphics[width=.95\linewidth]{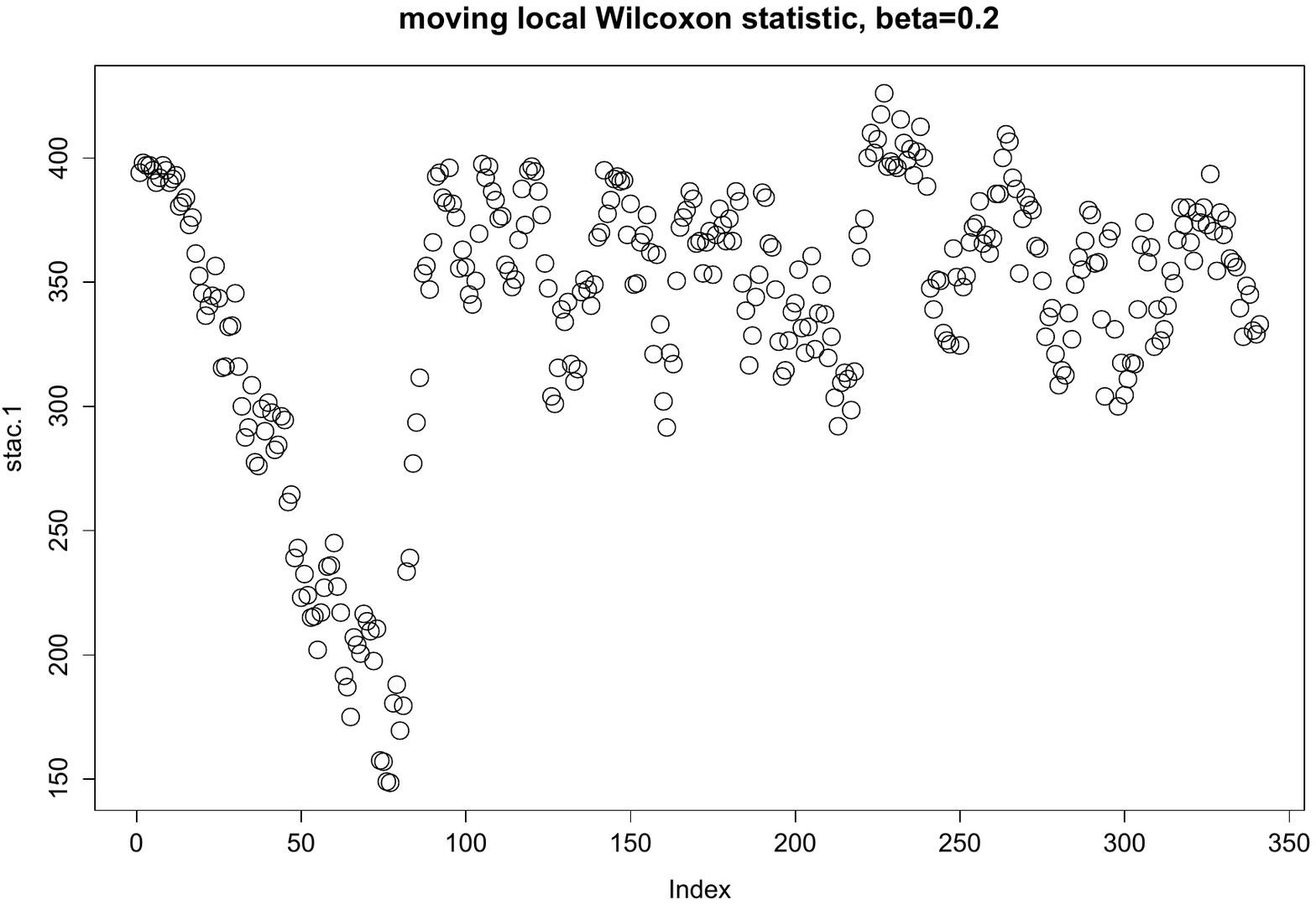}
\caption{Moving Wilcoxon statistic for numbers of users in service 1, $\beta=0.2$}
\label{fig16}
\end{minipage}
\end{figure}
Fig. 13 -- 16 show behavior of our proposal calculated from moving window for selected values of locality parameter $\beta$, and the reference sample consisted of the first 100 obs. One can notice a general tendency to stabilization of values of the statistic. The considered process seems to tend toward stationarity.

 \subsection{Empirical example 2 -- yield curves}
Our second empirical  example introduces FDA into modeling and predicting yield curves. Yield curves originate from the concept of risk free interest rate, i.e. theoretical price which is paid for investments in safe assets. In practice, however, risk-less instruments do not exist, the risk free rate is not directly observable and must be approximated by products traded on the market, like treasury bills, treasury and corporate bonds, inter bank lending rates, forward rate agreements or swaps etc.
From our point of view, yield curves are functions of time to maturity $\tau$. A change in yield curve shape is considered to be the sign of change of
 expectations and the sign for change in real business cycle phase.
Unfortunately, one cannot observe full functions' shape, since bonds and other interest rate derivatives have fixed dates of expiration. One should also mention here, that the detailed theory of shape of yield curve and factors affecting it are not fully developed.
Estimation of yield curve is usually done in two ways: in a non-parametric setting via linear or splines approximation and using  bootstrap techniques or using parametric approach \citep{Diebold}.\\
Since US economy is a precursor of changes in global economy, we focus our attention on US yield curve. For our study we use daily observed US yields from the period of 2000-01-03  to 2016-03-30 with maturities between 1 month and 30 years. In order to check, whether our test is able to detect financial crisis, this subset is divided into two parts:
\begin{enumerate}
\item  $X$ - before Lehman Brothers bankruptcy (2000-01-03 -2008-09-14),
\item $Y$ - since the beginning of sub-prime crisis in 2008-09-15 till present.
\end{enumerate}
Both subsets are converted into FDA objects and described in the Fourier basis (see fig. 17).
\begin{figure}
\centering
\includegraphics[width=.9\linewidth]{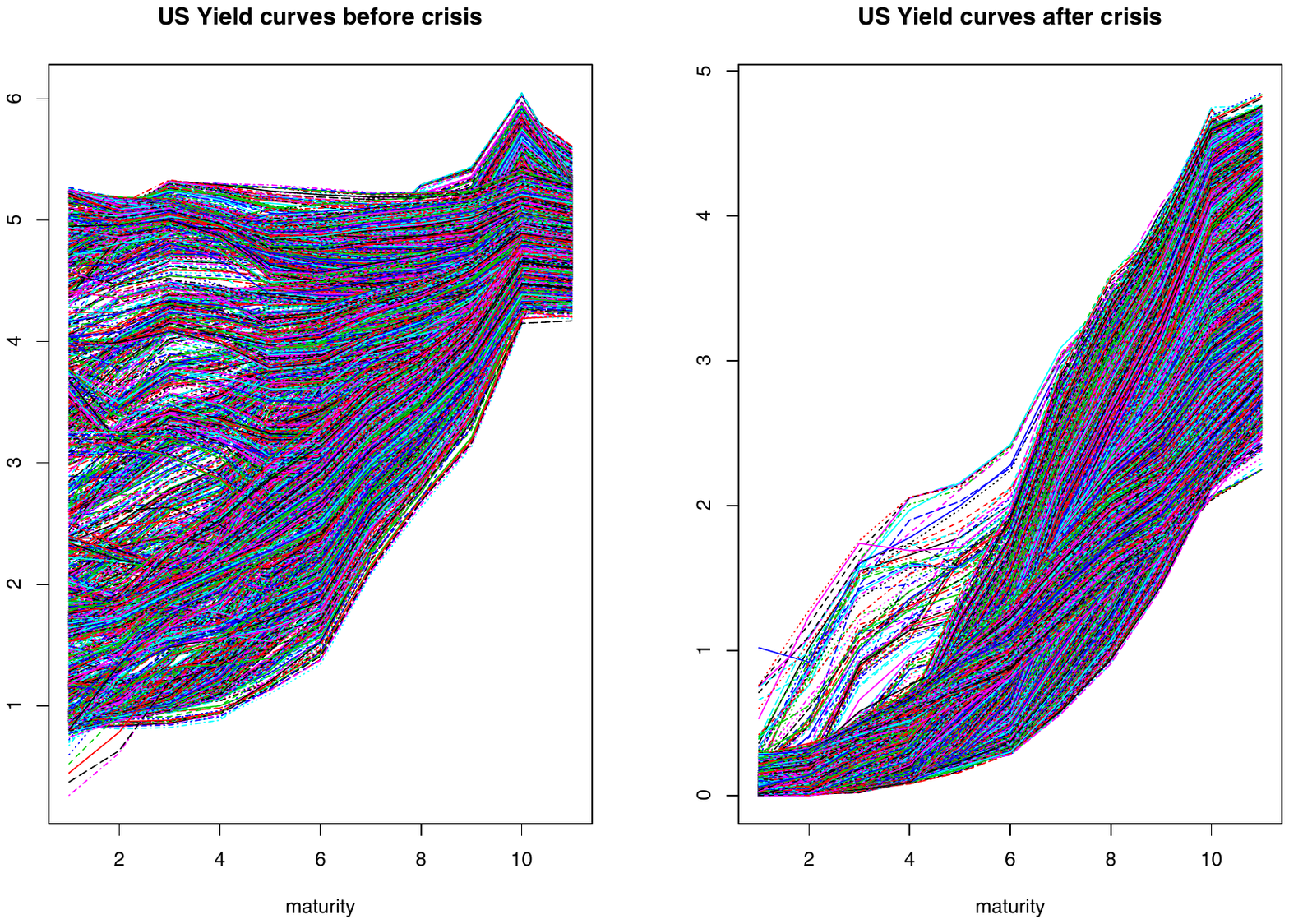}
\caption{Bundles of yield curves before and after crisis represented as functional object in Fourier basis}
\label{fig17}
\end{figure}
In fig. \ref{fig17} one can clearly see, that the  resultant shape (i.a. slopes and curvatures) for both bundles - before and after crisis are different, as they should, due to the change of business cycle phase in global economy.
The next step of the procedure requires the estimation of cGBD for both subsets. The functional boxplot in fig. 18 displays the median curve (the deepest location), along with the selected $\alpha$ central regions. Any point beyond the highest value of $\alpha$ may be considered as an outlier. 
As shown in fig. \ref{fig18} the central tendency of the shape of yield curve before and after crisis is the same, while the shape and nature of outliers differ significantly.
\begin{figure}
\centering
\includegraphics[width=.9\linewidth]{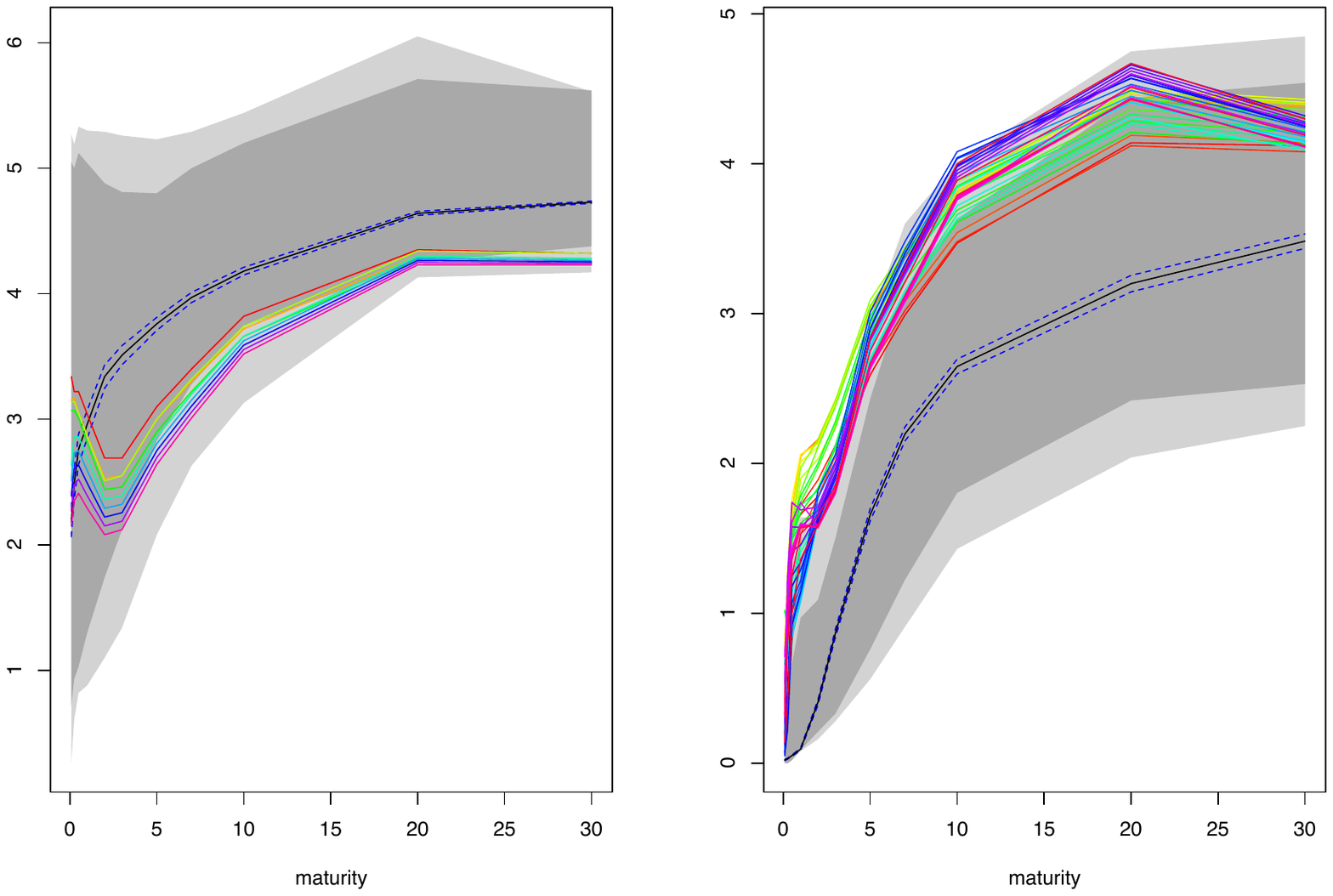}
\caption{Functional boxplot for both subsets of yield curves}
\label{fig18}
\end{figure}
 
\begin{figure}
\centering
\includegraphics[width=.9\linewidth]{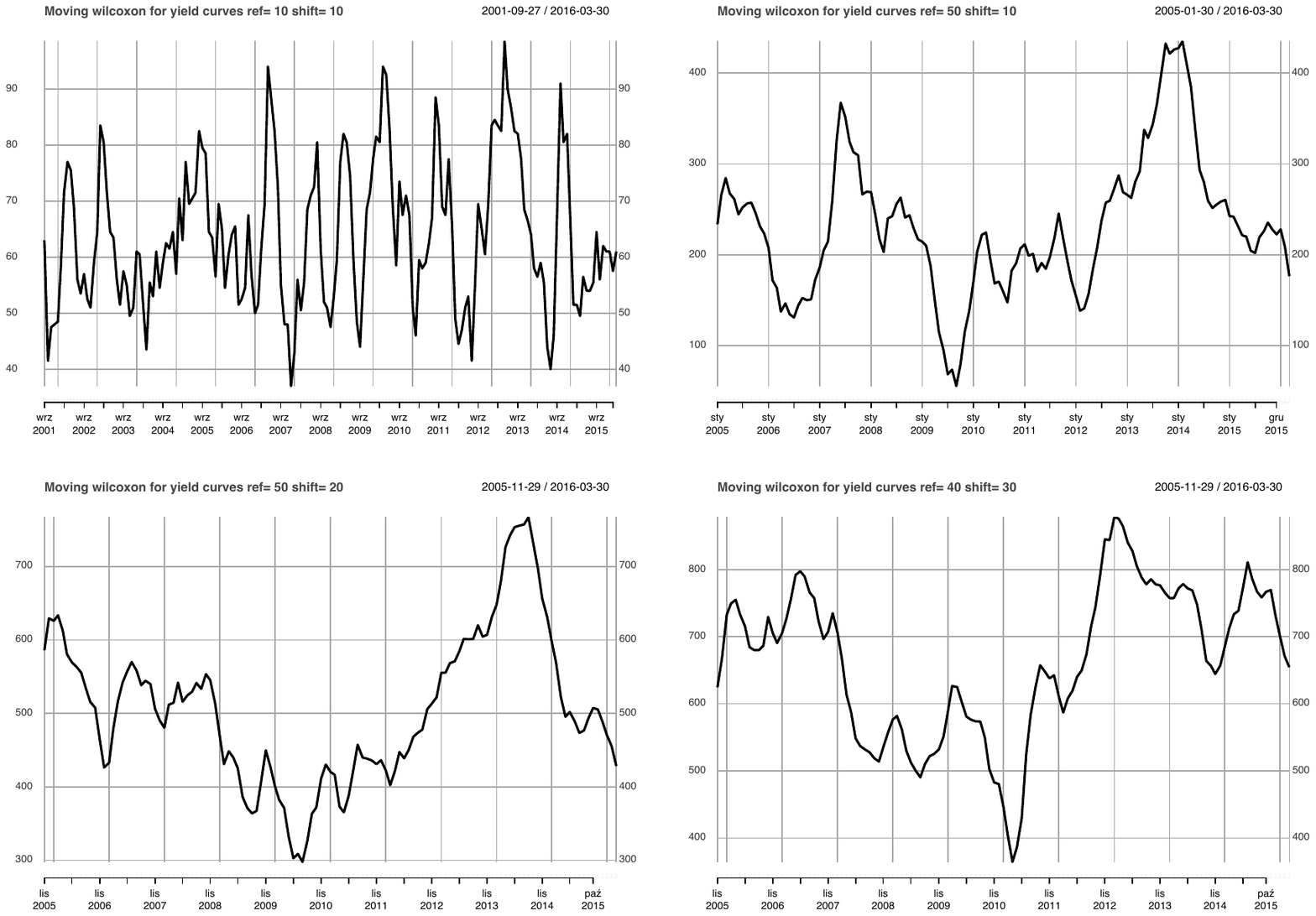}
\caption{Local Wilcoxon statistics for yield curves}
\label{fig19}
\end{figure}
In order to check whether our procedure is able to detect changes in the structure of functional time series, the final step of this simulation involved calculation of local Wilcoxon statistics for functional yield curve in the rolling window scheme (proposal 2). In this scheme we assume two windows of specified length - the reference window (or ref for short) has a fixed length and initially includes monthly functional yield curves data starting from Jan 2000, the second window of the same size is shifted by a fixed number of observations. The shift size and sample length are kept fixed. Results for windows of length 10, 50 and 40 observations shifted by 10, 20 or 30 points in time are presented in fig. 19. 
The obtained results clearly depends on window size. For relatively large windows and partially overlapping samples changes in the local Wilcoxon statistics are less volatile and one can relate them with phases in business cycle e.g. for windows of length 40 and 50 observations one can observe the regime change at the end of 2007 (or the begin of 2008) and changes between the end of 2009 and the begin of 2010 and in 2014. As yield curve is a predictor of the phase of business cycle, one can relate this sharp peaks to two crisis waves - sub-prime crisis and euro-zone debt crisis. Final peak can easily be related to the period of time when US economy entered a growth phase again. 
\section{Sensitivity analysis}
Classical one-dimensional Wilcoxon rank sum test effectively detects difference in location for family of logistic distributions. Multivariate tests induced by depths were proposed in \citep{Liu and Singh(1995)} and \citep{Liu}. Theoretical properties of multivariate Wilcoxon test (unbiasedness as well as its consistency were critically discussed in \citep{Jure}).
In our proposal ranks are induced by outlyingness relative to the local centrality characteristic. Observations are ranked from the closest to the local median to the furthest to the local median.
Relatively big or small values of the proposed statistic indicate differences in structure of  outlyingness (considered on a locality level $\beta$) and should lead us to rejecting a hypothesis of equality of distributions.

In the functional case, curves significantly differing may have the same depth and hence the same rank. On the other hand, different empirical depths indicate differences in the underlying distributions, because of the fact that under very mild conditions depths characterize multivariate distributions (if functional as well is still an opened question) \citep{Kong and Zuo(2010)}. However, the simulation studies lead us to a hypothesis that in a functional case we can expect a similar result: \emph{the corrected generalized band depth characterizes a distribution in a functional space or at least effectively describes its merit important features}.
Considering a reference sample and a moving window from a process we can use our proposal for detecting not only a structural change but also a departure from stationarity (represented by the reference sample).
Results of the simulations lead us to a conclusion that our proposal is at least qualitative robust in the Hampel sense (see \citep{RAND}).
 Small changes in null and alternative hypotheses do not significantly change size and power of our proposal. We considered distance in the input space in terms of a median of all distances for pairs of functions, where one function belong to assumed model and the second to a model representing a departure from assumptions. In the output space we considered euclidean distance between values of our test statistic. In theses terms small changes of input data lead to small changes of a decision process based on monitoring functional time series stationarity \citep{Hall}. A variety of possible outlyingness understandings in functional time series setting its worth to notice. It is possible to consider outliers in space of functions or outlyingness related to vertical point-wise contamination. Contamination may affect the reference sample or the working window. We considered functional outliers with respect to functional boxplot induced by the corrected generalized band depth.
Notice that our proposal is robust but not very robust (it copes with up to 10\% of contamination). It is robust to a moderate fraction of outliers or inliers (they lead to small change of ranking induced by depth) but sensitive to a time series regime change. The procedure may be used for data streams monitoring therefore \citep{Kos2016}.
In our opinion, alternative procedures for monitoring a homogeneity in functional time series are less robust to functional outliers than our proposal.
 
We can evaluate  the $"$size$"$ and the $"$power$"$ of our procedure in a similar manner as in \citep{Liu} and \citep{Jure}. A central issue in the analysis of functional data is to take into account the temporal dependencies between the functional observations. Due to this temporal dependence even the most elementary statistics became inaccurate.    
In this context resampling methodology, especially bootstrapping, turns out to be the only alternative. 
In order to obtain bootstrap p-values for our test, we propose to use a maximum entropy methodology proposed by \citet{Vinod} and used among other by \citet{Shang2016}. The meboot R  package together with DepthProc R package give the appropriate computational support.
  
\section{Summary}
The proposed procedure basing on moving local Wilcoxon statistic may effectively be used for detecting heterogenity in functional time series.
Simulation studies indicate that properties of our proposal depends on the Kolmogorov distance between functional medians in the distributions generating samples, one representing null hypothesis of stationarity and the second alternative representing a fixed departure from the stationarity. The locality parameter $\beta$ may be interpreted as a resolution or a sensitivity to details (e.g. local asymmetry) at which we monitor a process.

Merit properties of the proposed procedure strongly depend on the functional depth used (on which conditions we choose a center in a sample of functions compare e.g. \cite{Sguera,Nagy}. The conducted simulation studies as well as the studied empirical examples show a big potential of our proposal in a context of discrimination between the alternatives and in a a consequence in detecting  a structural change.
Implementations of the local Wilcoxon test and our proposal may be found in \emph{DepthProc} R package, which is available via CRAN servers. Note, that for detecting special kinds of nonstationarity it is possible to replace the local Wilcoxon statistic by means of local Kamat or Haga statistics (or other rank statistic).
Further theoretical properties of our proposal are still under our consideration and constitute part of our future work.

% %% \appendix

% %% \section{}
% %% \label{}

% %% References
% %%
% %% Following citation commands can be used in the body text:
% %% Usage of \citep is as follows:
% %%   \citep{key}          ==>>  [#]
% %%   \citep[chap. 2]{key} ==>>  [#, chap. 2]
% %%   \citept{key}         ==>>  Author [#]

% %% References with bibTeX database:

% %\bibliographystyle{model1-num-names}
% %\bibliography{sample.bib}

% %% Authors are advised to submit their bibtex database files. They are
% %% requested to list a bibtex style file in the manuscript if they do
% %% not want to use model1-num-names.bst.

\begin{acknowledgements}
JPR research has been partially supported by the AGH local grant no. 15.11.420.038
MS research has been partially supported by Cracow University of Economics local grants
no.045.WF.KRYF.01.2015.S.5045, no.161.WF.KRYF.02.2015.M.5161 and National Science Center Grant no. NCN.OPUS.2015.17.B.HS4.02708.
and DK  research by the CUE grant 048/WZ-KS/07/2016/S/6048.
\end{acknowledgements}

\end{document}